\documentclass[acmsmall]{acmart}

\setcopyright{rightsretained}
\acmJournal{TCPS}
\acmYear{2021} \acmVolume{1} \acmNumber{1} \acmArticle{1} \acmMonth{1} \acmPrice{}\acmDOI{10.1145/3485467}

\usepackage{ifthen}
\usepackage{xspace}
\usepackage[normalem]{ulem}
\usepackage{mathtools}
\usepackage{tikz}
\usepackage{pgfplots}
\usepackage{multirow}
\usetikzlibrary{positioning, arrows}
\usepackage{xcolor}
\usepackage{wrapfig}
\usepackage{capt-of}

\usepackage{etoolbox}
\AtBeginEnvironment{tabular}{\small}

\newcommand{\vs}{vs.\ }

\newcommand{\wrt}{w.r.t.\ }
\newcommand{\ie}{i.e.,\ }
\newcommand{\eg}{e.g.,\ }

\newcommand{\ieeestd}{IEEE~802.15.4\xspace}

\newcommand{\ctx}{CTX\xspace}

\newcommand{\pdr}{\ensuremath{\mathit{PDR}}\xspace}
\newcommand{\sdrnet}{\ensuremath{\overline{\mathit{SDR}}}\xspace}
\newcommand{\pdrnet}{\ensuremath{\overline{\pdr}}\xspace}
\newcommand{\dc}{\ensuremath{\mathit{DC}}\xspace}

\newcommand{\dcetc}{\ensuremath{\dc_\mathit{ETC}}\xspace}
\newcommand{\dcperiodic}{\ensuremath{\dc_\mathit{periodic}}\xspace}

\newcommand{\tbground}{\textsc{Hall}\xspace}
\newcommand{\tbdisi}{\textsc{Dept}\xspace}

\newcommand{\glossy}{Glossy\xspace}
\newcommand{\crystal}{Crystal\xspace}

\newcommand{\wcb}{\textsc{WCB}\xspace}
\newcommand{\etcpro}{\textsc{\wcb-E}\xspace}
\newcommand{\perpro}{\textsc{\wcb-P}\xspace}
\newcommand{\wcbe}{\etcpro}
\newcommand{\wcbp}{\perpro}

\newcommand{\SF}{\textsf{S}\xspace} 
\newcommand{\TF}{\textsf{T}\xspace} 
\newcommand{\AF}{\textsf{A}\xspace} 
\newcommand{\TAF}{\textsf{TA}\xspace} 
\newcommand{\EVF}{\textsf{EV}\xspace} 
\newcommand{\CTRF}{\textsf{CTRL}\xspace} 

\newcommand{\Twait}{\ensuremath{W}\xspace}

\newcommand{\WS}{\ensuremath{\Twait_{S}}\xspace}
\newcommand{\WEV}{\ensuremath{\Twait_{EV}}\xspace}
\newcommand{\WA}{\ensuremath{\Twait_{A}}\xspace}
\newcommand{\WT}{\ensuremath{\Twait_{T}}\xspace}
\newcommand{\WCTRL}{\ensuremath{\Twait_{CTRL}}\xspace}

\newcommand{\Ntx}{\ensuremath{N}\xspace}
\newcommand{\NtxS}{\ensuremath{\Ntx_{S}}\xspace}
\newcommand{\NtxEV}{\ensuremath{\Ntx_{EV}}\xspace}
\newcommand{\NtxT}{\ensuremath{\Ntx_{T}}\xspace}
\newcommand{\NtxA}{\ensuremath{\Ntx_{A}}\xspace}
\newcommand{\NtxCTRL}{\ensuremath{\Ntx_{CTRL}}\xspace}

\newcommand{\Etb}{\ensuremath{E}\xspace}
\newcommand{\Ctb}{\ensuremath{C}\xspace}
\newcommand{\Ttb}{\ensuremath{K}\xspace}
\newcommand{\Ttab}{\ensuremath{R}\xspace}

\newcommand{\Tfpp}{\ensuremath{T_\mathit{on,P}}\xspace}
\newcommand{\Tfep}{\ensuremath{T_\mathit{on,E}}\xspace}
\newcommand{\tfS}{\ensuremath{t_\mathit{on,S}}\xspace}
\newcommand{\tfEV}{\ensuremath{t_\mathit{on,EV}}\xspace}

\newcommand{\tfX}{\ensuremath{t_\mathit{on,X}}\xspace}
\newcommand{\pev}{\ensuremath{F_\mathit{ev}}\xspace}
\newcommand{\Ton}{\ensuremath{T_\mathit{on}}\xspace}

\newcommand{\epoch}{\ensuremath{T_\mathit{epoch}}\xspace}

\newcommand{\fakeparagraphnodot}[1]{\vspace{1mm}\noindent\textbf{#1}}
\newcommand{\fakeparagraphnospace}[1]{\vspace{1mm}\noindent\textbf{#1.}}
\newcommand{\fakeparagraph}[1]{\fakeparagraphnospace{#1}}

\usepackage{paralist}
\setdefaultenum{\em i)}{(a)}{1.}{A.}

\makeatletter
\renewcommand*{\p@section}{\S}
\renewcommand*{\p@subsection}{\S}
\renewcommand*{\p@subsubsection}{\S}
\makeatother

\usepackage[caption=false]{subfig}

\newboolean{authnotes}

\setboolean{authnotes}{false}

\ifthenelse{\boolean{authnotes}}
{
\newcommand{\tim}[1]{\footnote{{\bf Tim: #1}}}
\newcommand{\gp}[1]{\footnote{{\bf GP: #1}}}
\newcommand{\amy}[1]{\footnote{{\bf Amy: #1}}}
\newcommand{\matteo}[1]{\footnote{{\bf Matteo: #1}}}
\newcommand{\gabriel}[1]{\footnote{{\bf Gabriel: #1}}}
\newcommand{\mm}[1]{\footnote{{\bf MM: #1}}}

\newcommand{\rev}[1]{{\color{red} #1}}

\usepackage{draftwatermark}
\SetWatermarkLightness{0.85}
}
{
\newcommand{\tim}[1]{}
\newcommand{\gp}[1]{}
\newcommand{\amy}[1]{}
\newcommand{\matteo}[1]{}
\newcommand{\gabriel}[1]{}
\newcommand{\mm}[1]{}

\newcommand{\rev}[1]{#1} 
}

\def\d{\mathrm{d}}
\def\e{\mathrm{e}}
\def\R{\mathbb{R}}
\def\N{\mathbb{N}}
\newcommand*{\tran}{^{\mkern-1.5mu\mathsf{T}\!}}  
\def\norm[#1]{\left|#1\right|}  

\def\Linf{\mathcal{L}_{\infty}}

\def\xv{\boldsymbol{x}}
\def\yv{\boldsymbol{y}}
\def\uv{\boldsymbol{u}}
\def\wv{\boldsymbol{w}}
\def\xhat{\hat{\xv}}
\def\ev{\boldsymbol{e}}

\def\Am{\boldsymbol{A}}
\def\Bm{\boldsymbol{B}}
\def\Cm{\boldsymbol{C}}

\def\Em{\boldsymbol{E}}
\def\Km{\boldsymbol{K}}
\def\Qm{\boldsymbol{Q}}
\def\Rm{\boldsymbol{R}}
\def\Pm{\boldsymbol{P}}
\def\I{\mathbf{I}}
\def\Om{\mathbf{0}}
\def\Tm{\boldsymbol{T}}
\def\Mm{\boldsymbol{M}}
\def\Nm{\boldsymbol{N}}

\def\Is{\mathcal{I}}

\makeatletter
\DeclareRobustCommand
\myvdots{\vbox{\baselineskip4\p@ \lineskiplimit\z@
		\hbox{.}\hbox{.}\hbox{.}}}
\makeatother

\begin{document}

\newpage

\title{The Wireless Control Bus: Enabling Efficient Multi-hop Event-Triggered Control with Concurrent Transmissions}

\author{Matteo Trobinger}
\email{matteo.trobinger@unitn.it}
\affiliation{%
  \institution{University of Trento}
  \streetaddress{v. Sommarive 9}
  \city{Povo, Trento}
  \state{Italy}
  \country{Italy}
  \postcode{38122}
}

\author{Gabriel de Albuquerque Gleizer}
\email{G.deAlbuquerqueGleizer@tudelft.nl}
\affiliation{%
  \institution{Delft University of Technology}
  \streetaddress{Mekelweg 2}
  \city{Delft}
  \state{The Netherlands}
  \country{The Netherlands}
  \postcode{2628 CD}
}

\author{Timofei Istomin}
\email{timofei.istomin@unitn.it}
\affiliation{%
  \institution{University of Trento}
  \streetaddress{v. Sommarive 9}
  \city{Povo, Trento}
  \state{Italy}
  \country{Italy}
  \postcode{38122}
}

\author{Manuel Mazo Jr.}
\email{m.mazo@tudelft.nl}
\affiliation{%
  \institution{Delft University of Technology}
  \streetaddress{Mekelweg 2}
  \city{Delft}
  \state{The Netherlands}
  \country{The Netherlands}
  \postcode{2628 CD}
}

\author{Amy L. Murphy}
\email{murphy@fbk.eu}
\affiliation{%
  \institution{Bruno Kessler Foundation}
  \streetaddress{v. Sommarive 14}
  \city{Povo, Trento}
  \state{Italy}
  \country{Italy}
  \postcode{38122}
}

\author{Gian Pietro Picco}
\email{gianpietro.picco@unitn.it}
\affiliation{%
  \institution{University of Trento}
  \streetaddress{v. Sommarive 9}
  \city{Povo, Trento}
  \state{Italy}
  \country{Italy}
  \postcode{38122}
}

\authorsaddresses{}
\authorsaddresses{%
Authors' addresses: Matteo Trobinger, Timofei Istomin, Gian Pietro Picco, \{matteo.trobinger, timofei.istomin, gianpietro.picco\}@unitn.it, University of Trento, v. Sommarive 9, Trento, Italy, 38122; Gabriel de Albuquerque Gleizer, Manuel Mazo Jr., \{G.deAlbuquerqueGleizer, m.mazo\}@tudelft.nl, Delft University of Technology, Mekelweg 2, Delft, The Netherlands, 2628 CD; Amy L. Murphy, murphy@fbk.eu, Bruno Kessler Foundation, v. Sommarive 14, Trento, Italy, 38122.
}


\renewcommand{\shortauthors}{M. Trobinger, G. de Albuquerque Gleizer, M. Mazo Jr.,
                            A. L. Murphy and G. P. Picco}

\begin{abstract}
Event-triggered control (ETC) holds the potential to significantly
improve the efficiency of wireless networked control
systems. Unfortunately, its real-world impact has hitherto been
hampered by the lack of a network stack able to transfer its benefits
from theory to practice specifically by supporting the latency and
reliability requirements of the aperiodic communication ETC
induces. This is precisely the contribution of this paper.

Our \emph{Wireless Control Bus (\wcb)} exploits carefully orchestrated
network-wide floods of concurrent transmissions to minimize overhead
during quiescent, steady-state periods, and ensures timely and reliable
collection of sensor readings and dissemination of actuation commands
when an ETC triggering condition is violated. Using a cyber-physical
testbed emulating a water distribution system controlled over a
real-world multi-hop wireless network, we show that ETC over \wcb
achieves the same quality of periodic control at a fraction of the
energy costs, therefore unleashing and concretely demonstrating its
full potential for the first time.

 
\end{abstract}

\ccsdesc[500]{Networks~Network protocol design}
\ccsdesc[300]{Computer systems organization~Sensors and actuators}
\ccsdesc[300]{Information systems~Process control systems}

\keywords{event-based control, wireless sensor and actuator networks,
concurrent transmissions.}

\maketitle

 \section{Introduction}
\label{sec:intro}

As a result of the joint effort of academia and industry, low-power
wireless sensor networks (WSNs) are today a well-established
technology, proven to be very dependable and energy-efficient. In the
last twenty years, they have become the leading solution in a wide
domain of applications, including environmental
monitoring~\cite{CardellOliver2004}, wildlife
tracking~\cite{wildscope}, smart cities~\cite{tunnel}, and the
Internet of Things (IoT) at large~\cite{Pujolle2006}. This is due to
the high scalability and (re)placement flexibility, yielding lower
installation and maintenance costs, and to ever-improving computing
and communication features available on their untethered,
autonomously-powered, small hardware footprint.

\fakeparagraph{Low-power wireless networking for control: Challenges}
The benefits are so significant that low-power wireless networking is
now appealing also in traditionally wired domains like industrial
control~\cite{Khakpour2008, MarcoICCPS19}.
Nonetheless, although WSNs are widely adopted for \emph{monitoring},
their use for control and automation of plants and processes is still
very limited~\cite{ahlen2019toward}.
Key concerns hampering wider adoption are the reliability of
communication and the stability and magnitude of its latency.
Modern controllers depend on the reliable and timely communication of
relatively small data packets containing measurements and commands,
generated frequently at the sensors and controller. Guaranteeing these
properties is challenging in the large-scale, multi-hop scenarios that
are often the main reason for a wireless approach. Moreover, staple
applications for wireless control rely on battery-powered sensors,
which places energy efficiency in the limelight, as replacing batteries
is often costly or impractical. In this respect, it is well-known that
radio activity, both listening and transmitting,
is the main source of energy
consumption. Therefore, the design of low-power wireless protocol
stacks capable of minimizing communication without hampering control
performance is of utmost importance for the widespread adoption of
wireless control systems.

\fakeparagraphnodot{Event-triggered control: A missed opportunity?}
To facilitate the design of communications and simplify the control
performance analysis, most networked control systems (NCS), whether
wireless or wired, employ the classical periodic sampling of sensor
data and update of actuator commands. The choice of sampling period
involves a conservative, worst-case analysis of the closed-loop system
dynamics. However, this conservative design enters in direct conflict
with the objective of reducing energy consumption, enabled by the
low-power WSN operation and key to \emph{wireless} NCS (WNCS). To
address this limitation, aperiodic methods adapting to the dynamic
needs of the system have been investigated for a couple of decades
(see, e.g., \cite{aastrom2002comparison}).

A strong surge of interest began in 2007 with the systematic way of
designing aperiodic sampling proposed in~\cite{tabuada2007event},
currently known as \emph{event-triggered control (ETC)}, revolving
around the design of a \emph{triggering condition} that only depends
on sensor data. While this condition remains unsatisfied, a reference
Lyapunov function decays at a certain speed\footnote{Lyapunov
  functions are widely employed in stability and performance analysis
  and design of control systems. Informally, it can be seen as a
  mathematical generalization of the energy of a system: it is always
  positive, it grows with the magnitude of the states, and it is zero
  only at the desired equilibrium point. A decaying Lyapunov function
  implies that the system is approaching the equilibrium point. For an
  exposition, see, \eg~\cite{khalil2001nonlinear}.}; otherwise, as
soon as it is satisfied, sensor data is transmitted and control
commands are updated. This procedure guarantees a prescribed decay of
the Lyapunov function, serving as a certificate of performance for the
control system, while significantly reducing the need for
communication and, at least in principle, energy consumption.

Since then, many researchers embraced ETC and contributed to its
theoretical
foundations~\cite{wang2008event,girard2014dynamic,heemels2013periodic,mazo2011decentralized,tabuada2007event}.
However, its application is still problematic. Although ETC naturally
fosters resilience to communication delays~\cite{tabuada2007event},
this tolerance has its own limitations, and the latency of
communication imposes a limit on the achievable performance in terms
of convergence rate to an equilibrium. Therefore, minimizing delays
remains a critical goal for network stacks supporting ETC. Similar
comments hold for reliability, whose crucial role is exacerbated as
the entire network must timely and reliably react to the violation of
triggering conditions for ETC to operate properly.

Guaranteeing these and other properties with a proper network stack is
the most significant hampering factor to a wider adoption of
ETC. Although wireless implementations of ETC exist~\cite{Manuel_AWCS,
  van2016experimental, ramesh2013net, dolk2017platoon,
  kartakisfu2018commschemes}, these are limited to small-scale,
single-hop networks and exhibit poor reliability, high energy
consumption, large and unpredictable delays, or a combination thereof,
ultimately preventing the overall system to seize the energy savings
potentially enabled by ETC. This state of affairs is eloquently
summarized in a recent survey on wireless control
(\cite{marco-smart_manuf}, p.~22):
\begin{quote}
  ``While in the control community, many so-called event-triggered
  estimation and control approaches have been developed in the last
  two decades, it remains largely unclear whether and how these can be
  \emph{integrated} with the communication system and indeed result in
  demonstrable resource reallocation, savings, or other advantages for
  wireless systems in practice.''
\end{quote}

\fakeparagraph{A wireless control bus for ETC} In this paper, we
answer this question by providing a full-fledged network stack
operating in conjunction with ETC, therefore unlocking its remarkable
potential for energy savings hitherto hampered by the lack of
appropriate communication support.

Our approach exploits \emph{concurrent transmissions} (CTX) on the
same radio channel, a technique popularized by \glossy~\cite{glossy}
that has proven a very effective building block for protocol
design. Several protocols embraced this technique\footnote{Some
  authors use the label \emph{synchronous transmissions} instead, with
  equivalent meaning in this context.}, pushing the
envelope of what can be achieved by \ieeestd and, recently, other
low-power wireless radios including BLE, UWB, and
LoRa~\cite{ctx-survey}.

CTX-based protocols achieve at once very low latency, high
reliability, low energy consumption, and accurate time
synchronization. Based on efficient network-wide floods, they require
neither a MAC nor a routing layer, and their performance is largely
unaffected by changes in the topology induced, e.g., by node and link
failures.
This is a significant departure from conventional techniques (e.g.,
WirelessHART~\cite{wirelessHART}, ISA100.11.a~\cite{ISA},
6TiSCH~\cite{6TiSCH}) that mitigate the packet losses and missed
deadlines induced by network vagaries with continuous, high-overhead
topology maintenance.

Instead, CTX-based protocols allow for the communication medium to be
abstracted into a globally shared bus~\cite{lwb}; application data is
broadcast to the entire network and can therefore be read by each
node. In our context, this makes centralized control more
appealing and even efficient than decentralized and distributed
alternatives.
Centralized controllers are
generally easier to design and provide better performance than
controllers accounting for network topology constraints; further, in
the specific case of ETC they usually lead to fewer events being
triggered. Unfortunately, the use of CTX in control is hitherto
largely unexplored, apart from few recent
exceptions~\cite{MarcoICCPS19, MarcoTCPS, MarcoSelf-triggered} that
however focus on periodic and self-triggered sampling rather than ETC.

\fakeparagraph{Methodology and contributions} To achieve the
remarkable potential benefits of CTX-based communications in ETC,
co-design is fundamental. The control algorithm must work hand-in-hand
with the underlying network stack to seize opportunities to reduce the
radio active time while ensuring the timeliness and reliability key to
control performance. In ETC, control update times are not defined a
priori; sensors decide on-the-fly whether to send updated readings
based on their triggering condition.  This \emph{in theory} reduces
communication \wrt classic control approaches; \emph{in practice}, it
must be supported by a network stack capable of
\begin{inparaenum}
\item minimizing network overhead during the control idle times, and
\item promptly react to triggered events by ensuring timely and
  reliable collection of sensor readings at the controller and
  dissemination of updated actuation commands.
\end{inparaenum}

We address these challenges with the \emph{Wireless Control Bus
  (\wcb)}, a novel protocol that, to the best of our knowledge, is the
first supporting multi-hop communication for ETC, and does so
efficiently and reliably. We first summarize the technical foundations
of ETC and, motivated by the co-design of control and communication in
\wcb, put forth a side contribution further reducing communication via
rejection of step-disturbances (\ref{sec:etcprelim}). We then
illustrate how the design of \wcb (\ref{sec:design}) exploits \ctx to
meet the above requirements of ETC \wrt latency, reliability, and
energy efficiency. Moreover, we present a \wcb variant that can
easily accommodate conventional periodic strategies, endowing them
with similarly unprecedented performance and ultimately fostering a
holistic approach to control design enabled by a \emph{single} network
stack.

We demonstrate the effectiveness of our solutions via a
water-irrigation system (WIS) test case,
for which we define an ETC-based control strategy
(\ref{sec:testcase}). A WIS typically extends for kilometers, likely
requiring multi-hop communication, in turn demanding complex
decentralized or distributed control strategies, as
in~\cite{cantoni2007control}. In contrast, our combination of \wcb and
ETC enables a simpler centralized control, as we show experimentally.
In this respect, a realistic evaluation is a challenge per se, as we
are not aware of large-scale WIS testbeds. Small-scale ones, e.g., the
double-tank system~\cite{malmborg1997hybrid}, are widely adopted but
rely on a single-hop, star topology, unsuited to evaluate the
multi-hop systems envisioned for industrial wireless control and
targeted by this work.

We overcome these limitations with a secondary contribution: the design
of a cyber-physical testbed (\ref{sec:expsetup}) that adopts a real-time,
network-in-the-loop approach integrating
\begin{inparaenum}
\item a Simulink \emph{model} emulating the physical system, and
\item \emph{real} embedded devices acting as sensors, actuators,
  forwarders, and controller, executing our control and protocol stack
  and interacting only wirelessly.
\end{inparaenum}
We experiment with two distinct networks, where we analyze the
sensitivity of \wcb to its parameters (\ref{sec:protocolconfig}),
identify the configuration we use in our extensive experimental
campaign, and assess the impact of different scales and topologies on the
performance of our ETC system.

The experimental results (\ref{sec:eval}) demonstrate the
effectiveness of our approach. The quality of the control achieved by
ETC over \wcb is virtually the same as periodic sampling. However, it
comes at a fraction of communication costs; sample count is reduced by
$>$89\%, yielding a $>$62\% reduction in radio-on time \wrt periodic
control---far more than previously observed in the ETC
literature~\cite{Araujoself} in significantly more constrained
setups.
This confirms that \wcb not only provides a network stack, hitherto
missing, enabling ETC in multi-hop networks, but also effectively
translates the reduction of control traffic enabled by ETC into
corresponding savings in energy consumption.

The paper ends with a summary of related work (\ref{sec:relwork}) and
brief concluding remarks outlining opportunities for future work
(\ref{sec:ending}) on \wcb, which we intend to release publicly as
open source.


\section{Event-triggered Control}
\label{sec:etcprelim}

Event-triggered control (ETC) is a sampling strategy in which the
update of sensor data to feedback controllers and of control commands
to actuators is determined \emph{on-the-fly} by a \emph{triggering
  condition}. This is a drastic departure from time-triggered control,
which includes the classic periodic control.

In a nutshell, when something relevant happens on the state of a
dynamic system, the sensors communicate their most recent values to
the controller; otherwise, these values are held constant, and
actuators typically also hold their positions. Intuitively, data is
sampled \emph{only when needed}, reducing the communication induced by
control. In practice, determining when fresh data is needed is
somewhat involved and requires control theory to ensure stability and
good performance.

We formally describe ETC, including equations for a distributed
implementation suited to \ctx. In doing so, we also present two
contributions:
\begin{inparaenum}
\item a generalization of the decentralized ETC strategy
  in~\cite{mazo2011decentralized} to a broader class of triggering
  conditions and sensor node arrangements (\ref{sec:decent}), and
\item an adaptation of unperturbed ETC strategies to the problem of
  step disturbance rejection (\ref{ssec:stepdist}).
\end{inparaenum}

\subsection{Sample-and-hold control}\label{ssec:sohc}
	
Hereafter, we consider a linear time-invariant (LTI) system with
measurable states of the form
	\begin{equation}\label{eq:linplant}
          \dot{\xv}(t) = \Am\xv(t) + \Bm\uv(t) + \Em\wv(t),
	\end{equation}
	where $\xv(t) \in \R^n$ is the vector of states, $\uv(t) \in
        \R^m$ is the vector of control inputs, $\wv(t) \in \R^p$ is
        the vector of exogenous unmeasured disturbances, and $\Am,$
        $\Bm$, $\Em$ are known system matrices of appropriate
        dimensions. In this work, we assume that all states are
        measured by sensors. For digital implementation, we consider a
        state-feedback controller realized in a sample-and-hold
        fashion:
	\begin{equation}\label{eq:controller}
	\uv(t) = \Km\xhat(t),
	\end{equation}
        where $\Km$ is a control gain matrix to be designed, and
        $\xhat(t)$ is the sampled state, which satisfies, for a
        sequence of sampling times $\{t_i\}_{i\in\N}$,
	\begin{equation}\label{eq:sampleandhold}
	\xhat(t) = \xv(t_i), \forall t \in [t_i, t_{i+1}).
	\end{equation}
	We say that the obtained closed-loop system is \emph{globally
          exponentially stable} if, for every initial condition
        $\xv(0)$, all of its solutions satisfy
        $|\xv(t)| \leq M|\xv(0)|\e^{-\rho t}$ for some
        $0 \leq M < \infty$ and $\rho > 0$, where $\rho$ is called the
        \emph{decay rate} of the system.
	
	When using \emph{periodic sampling}, the sequence
        $\{t_i\}_{i\in\N}$ satisfies $t_i = ih$, for some designed
        sampling time $h$. In ETC, the sequence of sampling times is
        \emph{not} known a priori; instead, it is generated based on
        some designed \emph{triggering condition} dependent on the
        states. Although there is a vast literature on ETC, this
        section focuses on mechanisms enabling two important practical
        aspects for WNCS:
  \begin{enumerate}[1.]
  \item \emph{Triggering conditions can be checked periodically.}
    This is known as periodic ETC, or PETC \cite{heemels2013periodic}, 
    and it can achieve a control performance arbitrarily close to that 
    of classical ETC~\cite{postoyan2013periodic}. Periodic checking of
    triggering conditions allows for an efficient scheduling of
    sleep times, which we exploit in the design of \wcb (\ref{sec:design}).
    In contrast, classical ETC requires continuous monitoring of triggering 
    conditions, forcing sensors to be always active and preventing energy savings.
  \item \emph{Triggering conditions can be checked locally on
    the sensor nodes.} The alternative of checking them on
    the controller side would require sensors to send
    data to it periodically, which would eliminate any
    communication-related energy savings.
	\end{enumerate}

    We detail these two aspects next. Moreover, we note that
    our focus on LTI systems is mainly due to the test case we
    consider, which can be tackled as an LTI control problem
    (\ref{sec:water_irrigation}). PETC, decentralized ETC, and
    robust ETC for general non-linear systems have been
    addressed in, e.g.,~\cite{postoyan2013periodic},
    \cite{mazo2011decentralized}, and~\cite{liu2015small},
    respectively. Stability analysis in the non-linear case
    differs from the one presented in this paper, but the
    structure of the ETC mechanism is the same. Hence, the \wcb
    wireless protocol proposed here can be applied to non-linear
    or robust control problems without any
    changes.
	
	\subsection{Periodic event-triggered control}\label{ssec:petc}
	
	Using the framework of \cite{heemels2013periodic}, we define a
        periodic event-triggered state-feedback system as the one
        captured by \eqref{eq:linplant}--\eqref{eq:sampleandhold} with
        the triggering times satisfying
	\begin{equation}\label{eq:petctrig}
	t_{i+1} = \inf\left\{t=kh>t_i, k \in \N \,\middle|\,
	\begin{bmatrix}\xv(t) \\ \xhat(t)\end{bmatrix}\tran
	\Tm \begin{bmatrix}\xv(t) \\ \xhat(t)\end{bmatrix} > \epsilon^2 
	\right\},
	\end{equation}
	where $\Tm$ is a triggering matrix to be designed and
        $\epsilon$ is a design parameter whose value controls the size
        of the terminal set to which the system converges. When
        $\epsilon = 0$, the system converges and stabilizes at the
        desired equilibrium. A small $\epsilon>0$ increases the
        inter-sample times at the expense of stabilizing a set around
        the equilibrium, of size proportional to $\epsilon$.  When
        persistent external disturbances $\wv$ are present, one cannot
        stabilize the origin; setting $\epsilon>0$ is necessary to
        prevent excessive sampling precisely when the system is
        essentially under control, i.e., close to equilibrium.
	
	Several tools are available to verify the stability of the
        closed-loop system using a given triggering matrix $\Tm$.  We
        recall now one of the results
        from~\cite{heemels2013periodic}:
	\begin{theorem}[\cite{heemels2013periodic}, Theorem III.4]\label{thm:heemels}
          With $\epsilon=0$ and $w(t) \equiv 0$, the PETC system \eqref{eq:linplant}--\eqref{eq:petctrig} is
          globally exponentially stable (GES) with decay rate $\rho$
          if there exist symmetric matrices $\Pm_1, \Pm_2$, and
          scalars $\alpha_{ij}\geq0, \beta_{ij}\geq0,$ and
          $\kappa_i\geq0$, $i,j\in\{1, 2\}$, satisfying\footnote{For a
            symmetric matrix $\Am = \Am\tran$, we say that
            $\Am \succ \Om$ ($\Am \succeq \Om$) if it is
            positive-(semi)definite.}
		\begin{subequations}
			\begin{equation*}\label{eq:cond1}
			\e^{-2\rho h}\Pm_i - \Am_i\tran\Pm_j\Am_i + (-1)^i\alpha_{ij}\Tm + (-1)^j\beta_{ij}\Am_i\tran\Tm\Am_i\succeq\Om, \quad \forall i,j \in \{1, 2\},\quad\text{and}
			\end{equation*}
			\begin{equation*}\label{eq:cond2}
			\Pm_i + (-1)^i\kappa_i\Tm \succ \Om, \quad \forall i \in \{1, 2\},
			\end{equation*}
			\begin{equation*}
\text{where}\;\;  \Am_1 \coloneqq \begin{bmatrix}\Am+\Bm\Km & \Om \\ \I & \Om\end{bmatrix}, \quad \Am_2 \coloneqq \begin{bmatrix}\Am & \Bm\Km \\ \Om & \I\end{bmatrix}. 			
			\end{equation*}
		\end{subequations}
              \end{theorem}

              We use this result in our test case (\ref{sec:testcase})
              to design appropriate triggering conditions, i.e., a
              matrix $\Tm$ that guarantees appropriate control
              performance for a given sampling time $h$.

	\subsection{Distributed event-triggered conditions}\label{sec:decent}
	The triggering condition in \eqref{eq:petctrig} is, in its
        most general form, a centralized one, i.e., all states are
        needed to determine when to sample. However, when sensors are
        remotely located \wrt each other, this approach becomes
        impractical. Fortunately, decentralized triggering conditions
        exist that address this issue. Here we focus on the strategy
        proposed in~\cite{mazo2011decentralized}, consisting of three
        key steps posing corresponding requirements on the network
        stack supporting control:
	\begin{enumerate}[1.]
	\item Each sensor has its own triggering condition, which can
          trigger a controller update \emph{independently} of readings
          from other sensors.
	\item Upon one sensor triggering, \emph{all} others must
          transmit their up-to-date readings to the controller.
	\item Finally, the controller updates its control command and
          sends it to the actuators.
	\end{enumerate}
	
        The following type of triggering condition is used as a
        starting point in~\cite{mazo2011decentralized}:
	\begin{equation}\label{eq:tabuadatrig}
	\norm[\xv(t)-\xhat(t)] > \sigma\norm[\xv(t)],
	\end{equation}
	where $\sigma$ is a triggering parameter and $|\cdot|$ is the
        Euclidean norm. This condition, introduced by the seminal work
        in~\cite{tabuada2007event}, essentially compares the
        \emph{sampling error} $\xv(t)-\xhat(t)$ against the state
        values themselves; if the error is large enough, it is time to
        update the measurements at the controller.
	
	The main observation in~\cite{mazo2011decentralized} is
  that by rewriting~\eqref{eq:tabuadatrig} one obtains the implication:
	\begin{equation}\label{eq:decentralizedorig}
	\sum_{i=1}^n(x_i(t) - \hat{x}_i(t))^2 - \sigma^2x_i^2(t) > 0 \Rightarrow \bigvee_{i=1}^n \left((x_i(t) - \hat{x}_i(t))^2 - \sigma^2x_i^2(t) > \theta_i\right)
	\end{equation}
	as long as $\sum_{i=1}^n\theta_i = 0$ for $n$ state
        variables. This enables using each of the $i$-th conditions in
        the RHS of~\eqref{eq:decentralizedorig} independently at each
        sensor. The triggering parameters $\theta_i$ can be designed
        offline or adapted online. Hereafter, we focus on the former;
        details of their computation are found
        in~\cite{mazo2011decentralized}.

	Observe that~\eqref{eq:tabuadatrig} can be cast in the form
        of~\eqref{eq:petctrig} with
        $\Tm = \begin{bsmallmatrix} \!(1-\sigma^2)\I & \,-\I \\ -\I &
          \I \end{bsmallmatrix}$ and $\epsilon = 0$.  Thus, a simple
        generalization of the approach described above is possible, to
        include a larger class of triggering conditions of the form
        \eqref{eq:petctrig}, where more parameters (i.e., all elements
        of $\Tm$) than simply $\sigma$ are to be designed. This
        introduces additional design flexibility for the triggering
        conditions, which can be used to further reduce the amount of
        communication triggered by the system.

        First, denote the sampling error $\ev \coloneqq \xhat-\xv$.
        Assume $q \leq n$ sensor nodes, each measuring one or more
        state variables, and denote by $\Is_j \subseteq \{1,2,...,n\}$
        those measured by node~$j$ with
        $\bigcap_{j=1}^q\Is_j = \emptyset$, i.e., each state variable
        is measured by only one node. Then, a triggering condition of
        the form
	\begin{equation}\label{eq:decentralizable}
	\ev(t)\tran\Mm\ev(t) - \xv(t)\tran\Nm\xv(t) > \epsilon^2,
	\end{equation}
	is decentralizable if the triggering matrices $\Mm = \Mm\tran$
        and $\Nm = \Nm\tran$ have the following
        structure:
        an element
        $M_{ii'}$ ($N_{ii'}$) is nonzero if and only if $i$ and $i'$
        belong to the same set $\Is_j$ for some sensor node~$j$. Then,
        denoting by $\xv_j, \ev_j, \Mm_j$ and $\Nm_j$ the subvectors
        and submatrices containing the rows and columns $\Is_j$ of
        $\xv, \ev, \Mm$ and $\Nm$, we obtain
        that~\eqref{eq:decentralizable} implies:
	\begin{equation}\label{eq:decentralized}
	\bigvee_{j=1}^q(\ev_j(t)\tran\Mm_j\ev_j(t) - \xv_j(t)\tran\Nm_j\xv_j(t) > \theta_j),\qquad \text{with }\sum_{j=1}^q\theta_j = \epsilon^2.
	\end{equation}
	
	To make triggering as infrequent as possible, during design
        one may want to maximize some norm of $\Nm$ and minimize
        $\Mm$, so that the negative term in~\eqref{eq:decentralizable}
        dominates the inequality. Note that the triggering condition
        \eqref{eq:decentralizable} admits the form
        in~\eqref{eq:petctrig} with
        $\Tm = \begin{bsmallmatrix}\Mm-\Nm & -\Mm \\ -\Mm &
          \Mm\end{bsmallmatrix}$, therefore Theorem~\ref{thm:heemels}
        can be used to verify global exponential stability. This
        theorem can also be used to co-design, and optimize for sparse
        sampling, the matrices $\Pm_i$ and the triggering matrices
        $\Mm$ and $\Nm$; by fixing the values of
        $\kappa_i, \alpha_{ij}$, and $\beta_{ij}$, the problem becomes
        a linear matrix inequality (LMI) that can be easily solved
        with existing optimization software. To prevent the triggering
        condition from being repeatedly violated after the previous
        sample, when $\ev(t_i) = \Om$, $\Nm$ must be positive
        semidefinite.

	\subsection{The problem of disturbance rejection}\label{ssec:stepdist}
	
	The ETC mechanisms presented in this section are associated
        with the problem of stabilizing the origin, disregarding the
        effects of disturbances. Still, the presented triggering
        strategies also give disturbance attenuation properties in the
        case of linear systems. For example, sufficient conditions to
        verify a finite $\Linf$ gain are also present in
        \cite{heemels2013periodic}.
	
	In disturbance rejection problems, like the one we address in
        the WIS example on which we evaluate our solution, there is an
        important specificity: with the appropriate control design,
        one can ensure that a set of states (the control outputs
        $\yv(t) \in \R^p, \yv = \Cm\xv$) still converge to zero; the
        remaining states also converge, but to some unknown signal
        dependent on the disturbances (constant values in the case of
        step disturbances).
	
	If the objective is to stabilize the system to a given reference $\xv^*$, the general approach to event design is to perform a change of coordinates $\tilde{\xv} \coloneqq \xv - \xv^*$, which renders the problem again stabilizing $\tilde{\xv}$ to the origin. With this change of coordinates, note that the sampling error component does not change, i.e. $\hat{\ev} = \tilde{\xhat} - \tilde{\xv} = \xhat - \xv = \ev$. Condition \eqref{eq:decentralizable} becomes
	\begin{equation}\label{eq:decent_disturb}
	\ev(t)\tran\Mm\ev(t) - (\xv(t)-\xv^*)\tran\Nm(\xv(t)-\xv^*) > \epsilon^2.
	\end{equation}
	
	In the case of step disturbance rejection, some of the
        components of $\xv^*$ are unknown and vary depending on the
        disturbance. This makes it impossible to
        implement~\eqref{eq:decent_disturb} in its most general
        form. However, if one constrains the elements of $\Nm$
        associated with the unknown entries of $\xv^*$ to be zero,
        these terms do not appear in the equation, and the triggering
        condition is implementable regardless of the disturbance
        levels. Mathematically, the matrix on the second term
        of~\eqref{eq:decent_disturb} takes the form
        $(\Cm\tran\Cm)\Nm(\Cm\tran\Cm)$, and the triggering condition 
        can be implemented as
	\begin{equation}\label{eq:decent_disturb_y}
	\ev(t)\tran\Mm\ev(t) - \yv(t)\tran\Cm\tran\Nm\Cm\yv(t) > \epsilon^2,
	\end{equation}
	which can be decentralized to take the form in~\eqref{eq:decentralized}.
	To verify stability, one can use Theorem \ref{thm:heemels} with 
	$$ \Tm = \begin{bmatrix}\Mm - (\Cm\tran\Cm)\Nm(\Cm\tran\Cm) & -\Mm \\ -\Mm & \Mm\end{bmatrix}.$$
	

\section{Designing the Wireless Control Bus}
\label{sec:design}

The main focus of ETC is to avoid communication during steady-state,
while preserving correct and timely control outside of it. From a
network standpoint this means that
\begin{inparaenum}
\item when control traffic is absent, network overhead should be
  minimized; otherwise,
\item the collection of sensor readings at the controller and
  consequent dissemination of actuation commands should occur
  timely and reliably.
\end{inparaenum}
These requirements, already challenging when taken individually, are
even harder to fulfill when combined; a quiescent network, ideal to 
minimize consumption, is intrinsically at odds with a reactive and 
reliable one. It is therefore not surprising that a wireless network
stack efficiently supporting ETC is still missing, hampering the 
practical adoption of this control approach.

\wcb tackles this challenge by relying on concurrent transmissions
(\ref{sec:ctx}), whose peculiar properties are exploited to cater for
the specific needs of ETC (\ref{sec:etcpro}) and, within the same
protocol framework, also of traditional periodic control
(\ref{sec:perpro}).

\subsection{Concurrent Transmissions in a Nutshell}
\label{sec:ctx}

Conventional network protocols stagger transmissions to minimize
packet collisions. In contrast, protocols based on concurrent
transmissions (\ctx) exploit nodes transmitting \emph{at the same
  time}.

In \ieeestd, these protocols rely on two PHY-level
phenomena~\cite{ctx-tutorial, ctx-survey}. The so-called constructive
(or, more correctly, non-destructive) interference occurs when
\emph{identical} packets from multiple senders arrive at the receiver
with a time displacement $<$0.5~$\mu$s, the duration of a bit (chip)
in the transmitted chip sequence obtained by the direct-sequence
spread spectrum (DSSS) encoding of the original packet. In this case,
the signals are likely to mix non-destructively, and the packet is
successfully decoded.  The capture effect, instead, occurs even for
\emph{different} packets, as long as they arrive with a relative shift
$<$160~$\mu$s, the duration of the synchronization header;
one of the packets is likely received, depending on the density of
neighbors and their relative signal strength.

The effectiveness of \ctx has been demonstrated by the \glossy
system~\cite{glossy} that, originally designed for multi-hop time
synchronization, exploits the two phenomena above to achieve fast,
energy-efficient, and reliable network floods. The \emph{initiator}
begins a flood by broadcasting a packet. As the rest of the network is
assumed to be already listening on the channel, the packet is received
and immediately rebroadcast by neighbors, yielding \ctx. For
redundancy, each node retransmits the packet up to $N$ times. The
value of $N$ is key to determine the balance among reliability,
latency, and energy consumption. Similarly, the slot duration
must be short, to minimize the energy consumption due to listening, 
yet be long enough to accommodate all required packet transmissions.
Thanks to the massive concurrency, in practice the flood duration 
does \emph{not} depend on the number of nodes in the network but only
on the network radius, ensuring a latency---few milliseconds for few 
hops---very close to the theoretical minimum when using half-duplex
radios.

Since~\cite{glossy}, the popularity of \ctx increased dramatically,
leading to several low-power wireless systems significantly pushing
the performance boundary along several protocol dimensions, even in
PHY radio layers other than \ieeestd~\cite{ctx-tutorial,
ctx-survey} and in presence of harsh RF interference~\cite{competition17}.
These protocols typically exploit \glossy floods as
primitive building blocks, composing and scheduling them differently
in a distributed fashion, and exploiting either or both PHY-level
properties of \ctx depending on the protocol goals at hand. \wcb adopts a
similar approach, as described next.

\subsection{A Network Stack for Event-Triggered Control}
\label{sec:etcpro}

\fakeparagraph{Core concepts} Communication in \wcb is structured
around non-overlapping time \emph{slots}, each containing a separate
\glossy flood, potentially initiated by different nodes. The same
sequence of time slots repeats at all nodes with a fixed interval
called \emph{epoch}, characterized by a very short initial active
portion where communication occurs, and a much longer one where nodes
turn off their radio and remain in sleep mode.

This structure, common to many CTX-based systems, relies on the
accurate, network-wide time synchronization enabled by \glossy as part
of its operation, and effectively \emph{abstracts the multi-hop
  wireless network into a shared control bus with time-slotted
  access}. This simplifies significantly the development of the
overall control system by removing all the complexity typically
associated with multi-hop networks (e.g., at the MAC and routing
layers) and, at the same time, ensuring high determinism in terms of
latency and reliability---key for control design and performance.

Time slots can be
\begin{inparaenum}
\item dedicated to a single flood by one sender,
\item used by multiple senders concurrently flooding the same packet,
  or
\item by multiple senders flooding different packets competing in the
  same slot.
\end{inparaenum}
Although in all cases one packet is received with high probability,
experience with CTX-based systems shows that they offer decreasing
degrees of reliability (\ref{sec:ctx}).  \wcb balances the pros and
cons of each slot type depending on the target functionality,
described next.

\begin{figure}
  \centering
  \subfloat[Supporting event-triggered control: \etcpro.] {\label{fig:etc_protocol}\includegraphics[width=\textwidth]{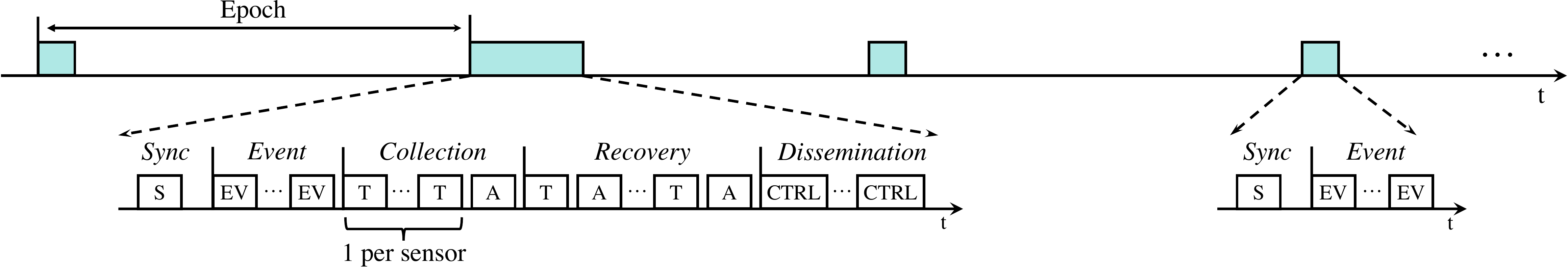}}\\
    \subfloat[Supporting periodic control: \perpro.] {\label{fig:per_protocol}\includegraphics[width=\textwidth]{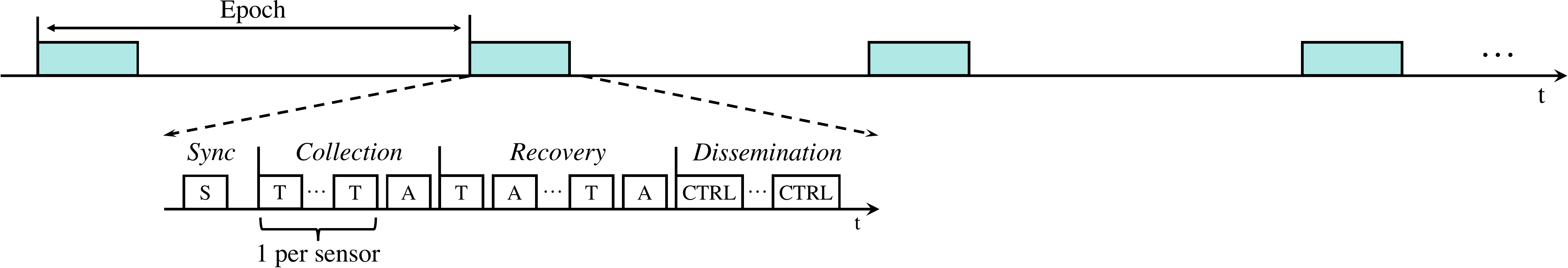}}
    \caption{The Wireless Control Bus, \wcb.}
    \label{img:protocol_struct}
\end{figure}

\fakeparagraph{Protocol phases} The active portion of a \wcb epoch is
structured in the following groups of functionally-related slots, or
\emph{phases} (Figure~\ref{img:protocol_struct}):
\begin{enumerate}[1.]
\item \emph{Synchronization.} CTX require tight time synchronization,
  which is also useful to establish a common time reference for
  control. However, prolonged sleeping periods---the main asset in
  reducing energy consumption---significantly increase clock
  drift. Therefore, as common in CTX-based protocols
  (\ref{sec:relwork}), each \wcb epoch begins with a \SF
  slot
  \rev{whose flood, initiated by the controller, contains the
    timestamp of packet transmission. This is used as a time reference
    by all other nodes that, by combining this information with the
    number of hops the packet has traveled, realign their local time
    reference that of the controller.}

\item \emph{Event.} This phase is key to efficient ETC support. After
  synchronization, each sensor node acquires its measurements and
  evaluates the triggering condition in~\eqref{eq:decent_disturb_y}
  (\ref{sec:etcprelim}). If this holds, a special and very short event
  notification packet---the same at all nodes---is flooded in one or
  more \EVF slots. Multiple events may be generated simultaneously at
  different nodes. However, due to the properties of CTX
  (\ref{sec:ctx}), this packet is received with very high reliability
  at all nodes, informing them \emph{at once} of the need to
  participate in the subsequent network-wide data collection (left
  schedule, Figure~\ref{fig:etc_protocol}). Otherwise, if no event is
  generated, the nodes can safely enter sleep for the remaining portion
  of the epoch (right schedule, Figure~\ref{fig:etc_protocol}).

\item \label{phase:collect} \emph{Collection.} Sensors report their
  readings as a sequence of \TF slots, each reserved to a sensor node
  performing an isolated flood. At the end, the \AF slot is reserved
  for an acknowledgment flood by the controller, containing a bitmap
  denoting which sensor packets have been successfully
  received. Thanks to the reliability of CTX, most of the times all
  reports are gathered, and all nodes can enter sleep until the
  dissemination phase (step~\ref{phase:disseminate}).
  
\item \emph{Recovery.} In the rare cases where a sensor node does not
  receive an acknowledgment or realizes that its packet is not
  confirmed in the bitmap, the node attempts retransmission in the
  subsequent \TF slot. Unlike collection, where each node transmits in
  a designated slot, during recovery unacknowledged sensors
  \emph{compete} in the same \TF slot with concurrent floods for their
  missed packets. 
  Again due to the properties of CTX (\ref{sec:ctx}), one of these
  packets reaches with high probability the controller, which updates
  the acknowledgment bitmap and floods it back in the \AF slot,
  effectively eliminating one of the competing nodes from the next
  \TAF slot pair.  This alternating sequence repeats until the
  controller acknowledges all packets, allowing nodes to safely enter
  sleep until dissemination, or a pre-defined number \Ttab of \TAF
  pairs is executed.

\item \label{phase:disseminate} \emph{Dissemination.} After collecting
  sensor readings, the controller generates the actuation commands. In
  the unlikely case where some readings are still missing after
  recovery, their values from the previous collection are employed by
  the controller.  This is the choice best aligning with the
  properties of ETC (\ref{sec:etcprelim}), although alternative ones can
  be easily integrated, if required. 
  Actuation commands are packed in a single packet and disseminated in
  one or more \CTRF slots by a controller-initiated flood; actuators
  apply the received commands upon their arrival. We \emph{always}
  include commands for \emph{all} actuators, even when their state is
  unchanged \wrt the previous dissemination, as this provides
  actuators with multiple chances to receive occasionally missed
  commands. Dissemination is the last phase of the epoch active
  portion; upon completion of the last \CTRF flood, the network
  automatically deactivates and all nodes enter sleep mode.
\end{enumerate}
\fakeparagraph{Ensuring reliability} Each phase exploits different
mechanisms to guarantee packet delivery. Recovery exploits an
acknowledgment slot \AF after a \TF slot, enabling competing nodes to
determine whether their packet has been received. This technique has
proven very effective~\cite{crystal} when the number of concurrent
transmitters is a priori unknown.
Nevertheless, in the collection phase it would double the number of
slots required and therefore latency and energy consumption. Instead,
we exploit a priori knowledge that \emph{all} sensors nodes must
transmit, and send a single, cumulative acknowledgment in the \AF slot
at the end of collection, itself triggering recovery only when needed.

\rev{Moreover, we note that a similar strategy could also lead to an
  alternative design of the recovery phase. The acknowledgment bitmap
  ending the collection phase provides nodes with enough global
  information to schedule their retransmissions back-to-back in
  dedicated slots, replied to by a single, final, collective
  acknowledgment. This scheme reduces the number of slots in the
  recovery phase and the contention in the \TF slots, and may be
  useful when several packets must be recovered at once. Nevertheless,
  it hinges on the correct reception of the acknowledgment bitmap,
  whose loss may be more common precisely in scenarios with several
  packet losses. Ultimately, these tradeoffs depend on the target
  environment; as our test environments
  (\ref{sec:protocolconfig},~\ref{sec:eval}) reveal very few lost
  packets, we use the simpler mechanism with competing retransmissions
  and individual acknowledgments.}

The mechanisms above are effective when packets must be delivered to a
\emph{single} node---the controller---that can signal their failed
receipt. However, they are impractical when packets must reliably
reach \emph{multiple} nodes, as in the event and dissemination
phases. In these cases, we exploit \emph{redundancy} as a simple yet
effective technique to increase reliability, and repeat the \EVF or
\CTRF multiple times. The number of repetitions is crucial, as it
governs the tradeoffs between reliability and energy consumption; we
analyze this parameter experimentally in~\ref{sec:protocolconfig}.

Finally, we exploit \emph{channel hopping} to further increase
resilience to interference, common in industrial scenarios but also in
indoor settings (e.g., due to WiFi) like those in our experiments
(\ref{sec:expsetup}). As \wcb nodes execute the same schedule in
lockstep, even during the dynamic recovery portion, the frequency
channel to be used in each slot can change following a globally-known
hopping sequence. \rev{This technique is considered state-of-the-art in the
context of CTX, as we further discuss in~\ref{sec:eval:nw}. Its
effectiveness towards interference resilience, without hampering
latency and energy consumption, has been thoroughly demonstrated
in~\cite{crystal2}, which directly inspires our design.}

\subsection{One Wireless Bus to Rule Them All: Periodic Control over
  \wcb}
\label{sec:perpro}

Our stated goal for the design of \wcb is to efficiently support
ETC. Nevertheless, our protocol can be easily tailored to periodic
control by regarding it as a special case of ETC in which the
triggering condition is violated during \emph{all} epochs. This
renders the dynamic and distributed coordination offered by the event
phase superfluous, leading to the schedule in
Figure~\ref{fig:per_protocol}. Hereafter, we refer to this specific
variant targeting periodic control as \wcbp whenever necessary to
distinguish it from the original protocol targeting ETC
(Figure~\ref{fig:etc_protocol}), itself referred to as \wcbe.

Although the modifications leading to \wcbp are simple, their impact
should not be underestimated. On one hand, the dedicated support
offered by \wcbe to ETC remains crucial. The active periods in \wcbp
are generally longer than in \wcbe, resulting in significantly less
energy-efficient communication,
as hinted at by the larger active portions of the former in
Figure~\ref{img:protocol_struct} and quantitatively shown in our
experimental evaluation (\ref{sec:eval}). On the other hand, due to
the specific application and control requirements, periodic control
may be preferable to ETC. In these cases, the efficiency and
performance offered by \wcbp over multi-hop networks is
unprecedented. Further, the ability to use the \emph{same} protocol
stack for both flavors of control, ETC and periodic, is a tremendous
asset. Not only it greatly reduces the complexity of control design
and implementation, but also fosters a holistic approach where the
selection of the best control strategy is driven solely by application
requirements rather than the lack of a suitable network stack.


\section{Test Case: A Water Irrigation System}
\label{sec:testcase}

To validate experimentally \wcb in a realistic scenario, we use a
water irrigation system (WIS) as our test case. A WIS is constituted
by a set of pools, often a few kilometers long, connected to one other
with controllable gates whose movement regulates the levels of each
pool, providing customers with a relatively constant supply. Without
communication between neighboring gates, each gate regulates the level
of the pool immediately downstream or upstream without knowledge of
what happens on the neighboring pools, in what is known as
decentralized control.
In~\cite{cantoni2007control} and~\cite{li2011stability}, it is noted
that decentralized control has several limitations that can waste
water due to spillovers. These references suggest the use of more
interconnected types of control such as centralized and distributed
control architectures, in which information from neighboring pools (or
all pools in the centralized case) is shared to improve control. With
distances on the order of kilometers to be covered and the typical
lack of existing infrastructure in these areas, WIS are one of the
prototypical applications of control over multi-hop wireless
networks.
	
Here we describe our test case, which builds on a real
scenario~\cite{li2011stability}. We then present the periodic
event-triggered control (PETC) design that is the basis of our
experiments. It is not our intention in this paper to provide a
complete solution to WIS; instead, our goal is to use this example
as a proof-of-concept for the combination of ETC and \wcb presented
here. Therefore, we design a simple centralized state-feedback
controller that captures the essence of the centralized control
problem and allows us to showcase a centralized ETC solution over
wireless. Control solutions considering more practical design
criteria for WIS are available in,
e.g.,~\cite{cantoni2007control,li2008distributed}.

\subsection{System description and modeling}
\label{sec:water_irrigation}

\begin{figure}
	\begin{center}
		\begin{minipage}[t]{0.48\textwidth}
			\vspace{0pt}
			\includegraphics[width=\textwidth]{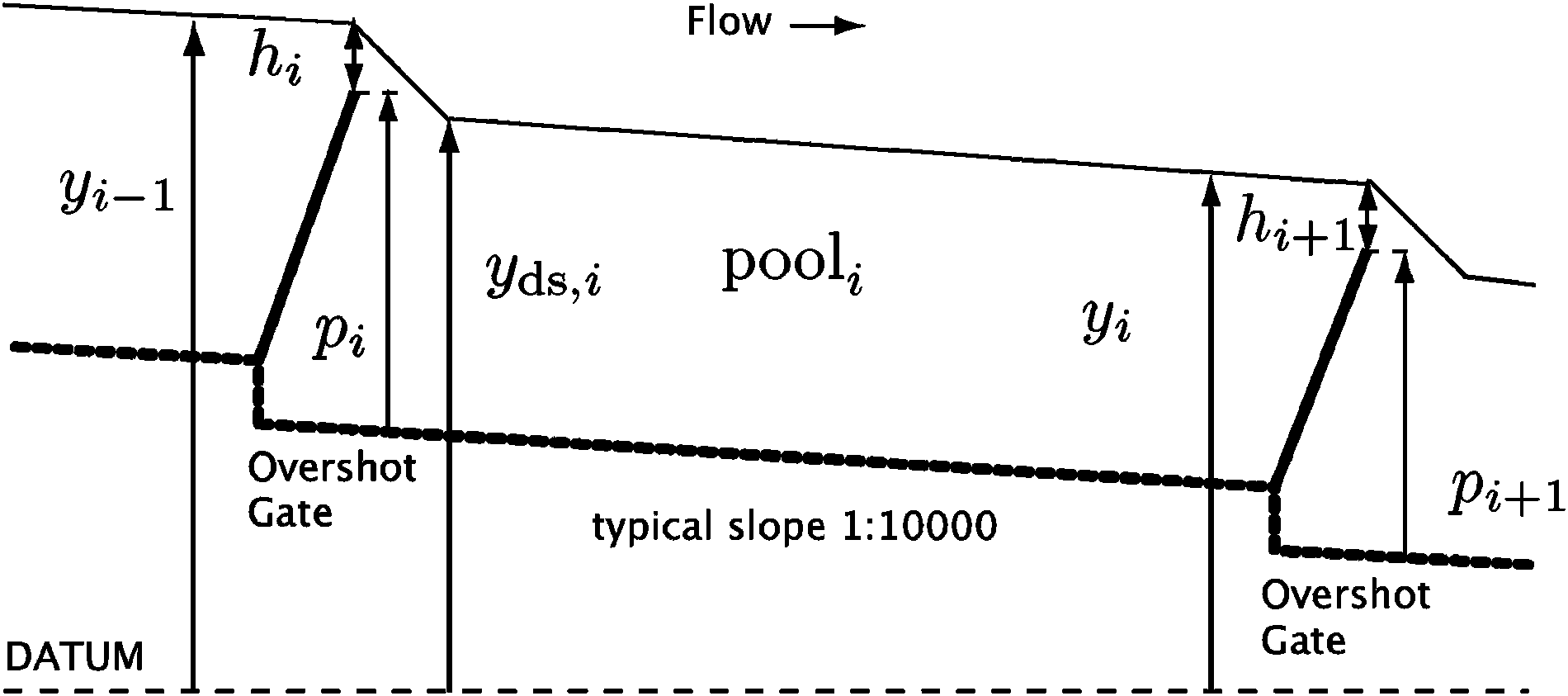}
			\caption{A section of an open-water channel with overshot gates (from \cite{cantoni2007control}).	\label{fig:pool}}     
		\end{minipage}\hfill
		\begin{minipage}[t]{0.48\textwidth}
			\vspace{0pt}
			\captionof{table}{\label{tab:water_params}Parameters of the WIS models \eqref{eq:pool_control} and \eqref{eq:pool_sim}: delay ($\tau_i$), surface area ($\alpha_i$), and dominant wave frequency ($\varphi_i$).}
			\begin{tabular}{llllll}
				\hline 
				Pool & 1 & 2 & 3 & 4 & 5 \\ 
				\hline 
				$\tau_i$ (min) & 4 & 2 & 4 & 4 & 6 \\  
				$\alpha_i$ (m\textsuperscript{2}) & 6492 & 2478 & 6084 & 5658 & 7650 \\ 
				$\varphi_i$ (rad/min) & 0.48 & 1.05 & 0.48 & 0.48 & 0.42 \\ 
				\hline 
			\end{tabular}
		\end{minipage}
	\end{center}
\end{figure}

In our test case, we consider a WIS composed of multiple pools
connected in series; a lateral view is depicted in
Figure~\ref{fig:pool}.  The control problem is to regulate the levels
of each pool to their setpoints by adjusting the position of the
gates. Opening the gates increases the flow from pool $i-1$ to pool
$i$, contributing to a reduction of level $y_{i-1}$ and an increase of
$y_i$. External off-take disturbances come mostly from end-users, and
typically occur downstream in each pool. The control objectives \wrt
level regulation are~\cite{cantoni2007control}:
\begin{inparaenum}
\item avoiding losses due to spillovers
\item keeping levels close to the setpoint to avoid oversupplying, and
\item preventing fluctuations occurring when dormant waves are
  excited.
\end{inparaenum}

Accurate models of open water dynamics are very complex. For control
design, we can use a simpler one capturing the first modes of wave
phenomena via the conservation of mass principle:
\begin{equation}\label{eq:pool_orig}
\pi_i\left(\frac{\d}{\d t}\right)y_i(t) = \gamma_i h_i^{3/2}(t-\tau_i) - \gamma_{i+1}h_{i+1}^{3/2}(t) - d_i(t),
\end{equation}
where $h_i$ is the relative height above gate $i$
(Figure~\ref{fig:pool}), $d_i$ is the total flow of off-take
disturbances, $\tau_i$ is the time for water to traverse the pool
length, and $\gamma_i$ is a parameter depending on the pool and gate
geometry. The model dynamics are captured by a polynomial
$\pi_i(\cdot)$: higher orders yield more accurate models. We assume
that the flow $u_i(t) = \gamma_i h_i^{3/2}(t)$ over gate $i$ can be
directly manipulated\footnote{An example of actuating device in this
  context is FlumeGate\textcopyright, by the company
  Rubicon~\cite{flumegate}.}, making~\eqref{eq:pool_orig} linear. For
control design, a first-order polynomial $\pi_i$
suffices~\cite{cantoni2007control,li2008distributed}
\begin{equation}\label{eq:pool_control}
\alpha_i \dot{y}_i(t) = u_i(t-\tau_i) - u_{i+1}(t) - d_i(t),
\end{equation}
where $\alpha_i$ is the pool surface area. However, this model is too
simplistic for simulation, an integral part of the experimental setup
(\ref{sec:expsetup}) supporting our combined evaluation of the control
and network layers (\ref{sec:eval}). Therefore, as
in~\cite{cantoni2007control}, we use a third-order polynomial
$\pi_i(\cdot)$ for the simulated plant:
\begin{equation}\label{eq:pool_sim}
\frac{\alpha_i}{\omega_{n,i}^2}(\dddot{y}_i(t) + 2\zeta_i\omega_{n,i}\ddot{y}(t) + \omega^2_{n,i}\dot{y}(t)) = u_i(t-\tau_i) - u_{i+1}(t) - d_i(t),
\end{equation}
where $\zeta_i$ and $\omega_{n,i}$ (satisfying $\varphi_i = \omega_{n,i}\sqrt{1-\zeta_i^2}$, for $\varphi_i$ the dominant wave frequency), represent the
first-mode wave damping ratio, and natural frequency of pool $i$ respectively.
In our test case, we consider a string of five pools representing a
section of a water channel in New South Wales, Australia. The
characteristics of this setup and related parameters
(Table~\ref{tab:water_params}) are found
in~\cite{li2011stability}. Moreover, we set the additional parameter
$\zeta_i = 0.0151$ for all $i$, as in~\cite{weyer2001system}.

\subsection{Event-triggered control design}
\label{sec:wis:control}

For ETC design, we apply the principle of separation of concerns
between control design and cyber-physical implementation. The
controller is designed as a continuous-time controller, for which many
methods are available. Then, a sampled-data implementation \rev{based
on PETC}
is devised, which must consider the imperfections
of the communication channel to retain some given performance
specifications. This prevents changes (e.g., in network technology,
topology, nodes, etc.) from requiring a complete redesign of the
controller. In our case, this is achieved with the following design
procedure:
\begin{enumerate}[1.]
	\item design a centralized state-feedback controller that rejects step disturbances; 
	\item select the sampling time $h$ for monitoring and
          event-checking; and
	\item design the distributed event-triggering parameters
          $\Mm_j, \Nm_j, \theta_j$ that achieve similar performance to
          the continuous-time controller (\ref{sec:etcprelim}).
\end{enumerate} 
	
To design a centralized ETC for the WIS in~\ref{sec:water_irrigation},
we need a state-space description of the system
in~\eqref{eq:pool_control}.
To this end, we replace the time-delay by its Pad\'{e} approximation of
order $(1,1)$, as in~\cite{cantoni2007control}, and extend the model with states $x_{3,i}$ integrating $y_i$, to enable rejection of persistent off-take disturbances by the controller.
A state-space representation of the resulting model is given by:
\begin{equation}\label{eq:sspoolint}
\dot{x}_{1,i} = -\frac{1}{\tau_i}{x}_{2,i} - \frac{1}{\alpha_i}(u_i + u_{i+1} + d_i), \qquad
\dot{x}_{2,i} = -\frac{2}{\tau_i}{x}_{2,i} + \frac{4}{\alpha_i}u_i, \qquad
\dot{x}_{3,i} = x_{1,i},
\end{equation}
where $x_{1,i} \coloneqq y_i$, $x_{2,i}$ can be
regarded\footnote{Alternatively, it can be viewed as the {Pad\'{e}}
approximant of the Smith predictor for the subsystem
$\alpha_i\dot{x}_{2i} = u_i(t-\tau_i).$} as a low-pass filter on the
flow $u_i$, and $u_6(t)=0,\;\forall\,t$, i.e., there is no controlled gate at the downstream side of the last pool. The variables $x_{2,i}$ and $x_{3,i}$ can be locally computed at the flow and height measurement nodes, respectively.

With this model, one can use standard state-space methods for control
design. For our test case, we designed a linear-quadratic regulator
(LQR) using diagonal weight matrices $\Qm$ and $\Rm$, with $\Rm = \I$
and $\Qm$ with diagonal entries (1250, 1250, 2500, 5000, 7500) for
$x_{1,i}$, 0 for $x_{2,i}$, and (1.25, 1.25, 2.5, 5, 7.5) for
$x_{3,i}$. These values were tuned to achieve a uniform convergence
across pools, a trade-off between speed of the state convergence and
magnitude of control action, and robustness w.r.t.~the natural
frequency of oscillation of the pools. The GES decay rate
(Theorem~\ref{thm:heemels}) of the continuous-time closed loop system
is $\rho$=0.007
min$^{-1}$.

\begin{figure}
	\begin{minipage}{0.48\textwidth}
		\begin{center}
			\includegraphics{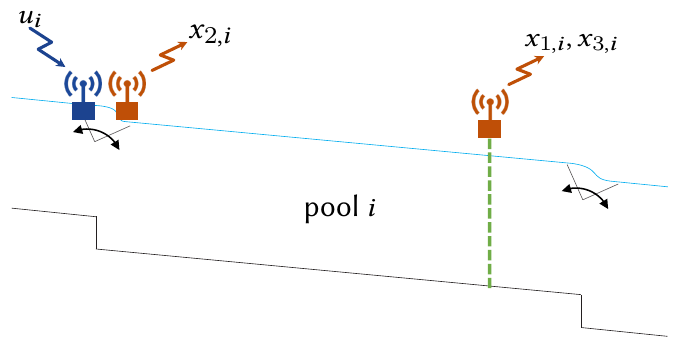}
			\vspace{-2em}
			\caption{\label{fig:pool_radio} Data
                          communicated to/from nodes at pool $i$. The
                          dashed green line denotes a height
                          measurement sensor, while L-shaped gray
                          elements denote gates with flow control and
                          measurement capabilities.}
		\end{center}
	\end{minipage}\hfill
	\begin{minipage}{0.48\textwidth}
		\begin{center}
			\includegraphics{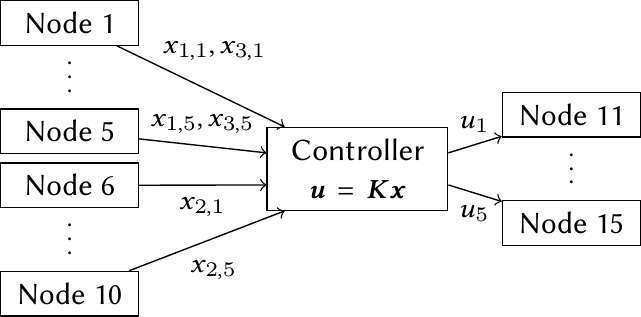}
			\caption{\label{fig:blockdiagram} Control data
                          diagram for the 5-pool system. Each of
                          nodes~6--10 is co-located with nodes 11--15,
                          respectively; therefore, they can be hosted
                          by the same physical device.}
		\end{center}
	\end{minipage}
\end{figure}

Figure~\ref{fig:pool_radio} illustrates how control data is
communicated wirelessly. The height sensor node also performs the
integration locally to compute $x_{3,i}$. The gate has one node to
receive control inputs $u_i$ and one to compute the filtered flow
value $x_{2,i}$ and send it to the controller. For the 5-pool system
we consider, a total of 10~sensor and 5~actuator nodes are used. The
height setpoints are assumed to be locally available to the height
device; hereafter, $x_{1,i} = y_i - y_i^*$, i.e., control regulates
deviations of height \wrt its setpoint, assumed to be set constant
throughout the experiment.  Figure~\ref{fig:blockdiagram} shows a
block diagram for the complete control system; note how the controller
is a separate node.

We choose the fundamental sampling period $h=1$~min as
in~\cite{weyer2001system}, where this value is used for short pools up
to 3200~m, as in our setup.
As for ETC, we solve iteratively the LMIs in Theorem~\ref{thm:heemels}
to find matrices $\Mm_j$ and $\Nm_j$ achieving a high sampling
performance (\ref{sec:decent}). The triggering parameters $\theta_j$
are tuned to further improve the latter in a trade-off with
steady-state error, for which a magnitude of 1~cm is deemed
acceptable. Figure~\ref{fig:ourmatrices} shows the values of $\Mm_j$,
$\Nm_j$, $\theta_j$. Nodes 1--5 represent height sensors, with
matrices partitioned according to
$[\begin{matrix}x_{1,j} \; x_{3,j}\end{matrix}]$, while nodes 6--10
represent filtered flow ($x_{2,j}$) sensors. The resulting decay rate,
satisfying Theorem~\ref{thm:heemels}, is $\rho$=0.006~min$^{-1}$.

\begin{figure*}[!t]
\renewcommand*{\arraycolsep}{1pt}
{\footnotesize
\begin{alignat*}{9}
\Mm_1 &=\!\begin{bmatrix}0.621 & 0.0030 \\ 0.003 & 0.0001 \end{bmatrix}\!, &
\!\Mm_2 &=\!\begin{bmatrix}0.414 & 0.003 \\ 0.003 & 0.0002\end{bmatrix}\!, &
\!\Mm_3 &=\!\begin{bmatrix}1.854 & -0.083 \\ -0.083 & 0.13\end{bmatrix}\!, &
\!\Mm_4 &=\!\begin{bmatrix}2.48 & 0.012 \\ 0.012 & 0.001\end{bmatrix}\!, &
\!\Mm_5 &=\!\begin{bmatrix}7.639 & 0.027 \\ 0.027 & 0.006\end{bmatrix}\!,\\
\Mm_6 &= 0.1147, &
\Mm_7 &= 0.0841, &
\Mm_8 &= 0.2337, &
\Mm_9 &= 0.5352, &
\Mm_{10} &= 1.4786,\\
\Nm_1 &= \begin{bmatrix}2.5\times 10^{-8} & 0 \\ 0 & 0 \end{bmatrix}, &
\Nm_2 &= \begin{bmatrix}0.0503 & 0 \\ 0 & 0 \end{bmatrix}, &
\Nm_3 &= \begin{bmatrix}1.2\times 10^{-8} & 0 \\ 0 & 0 \end{bmatrix}, &
\Nm_4 &= \begin{bmatrix}10^{-6} & 0 \\ 0 & 0 \end{bmatrix}, &
\Nm_5 &= \begin{bmatrix}0.9497 & 0 \\ 0 & 0 \end{bmatrix}, \\
\Nm_6 &= 0, &
\Nm_7 &= 0, &
\Nm_8 &= 0, &
\Nm_9 &= 0, &
\Nm_{10} &= 0,\\
\theta_1\! &= 0.415, \ \ \theta_2\! =0.24, & \theta_3\! &=0.987, \ \ \theta_4\! =1.18, & \theta_5\! &=2.15, \ \ \theta_j=9, \forall j\in\{&6,...&,10\}, & & 
\end{alignat*}
}
\vspace*{-6mm}
\caption{Triggering parameters applied in the test case.
  \label{fig:ourmatrices}}
\end{figure*}


\section{A Cyber-Physical Experimental Testbed}
\label{sec:expsetup}

A widely-adopted methodology for evaluating WNCS relies on small-scale
laboratory setups mimicking industrial process control loops, e.g.,
the double-tank system~\cite{Araujoself, Manuel_AWCS}. This approach
tests the ability to control \emph{real} physical processes, but often
relies on single-hop networks, neglecting key networking aspects
(e.g., packet delays and losses) which WCB instead explicitly
addresses.

\begin{wrapfigure}{r}{6.9cm}
  \vspace*{-4mm}
  \centering
  \subfloat[Architecture.] 
  {\label{fig:architecture}
  \includegraphics[scale=0.5]{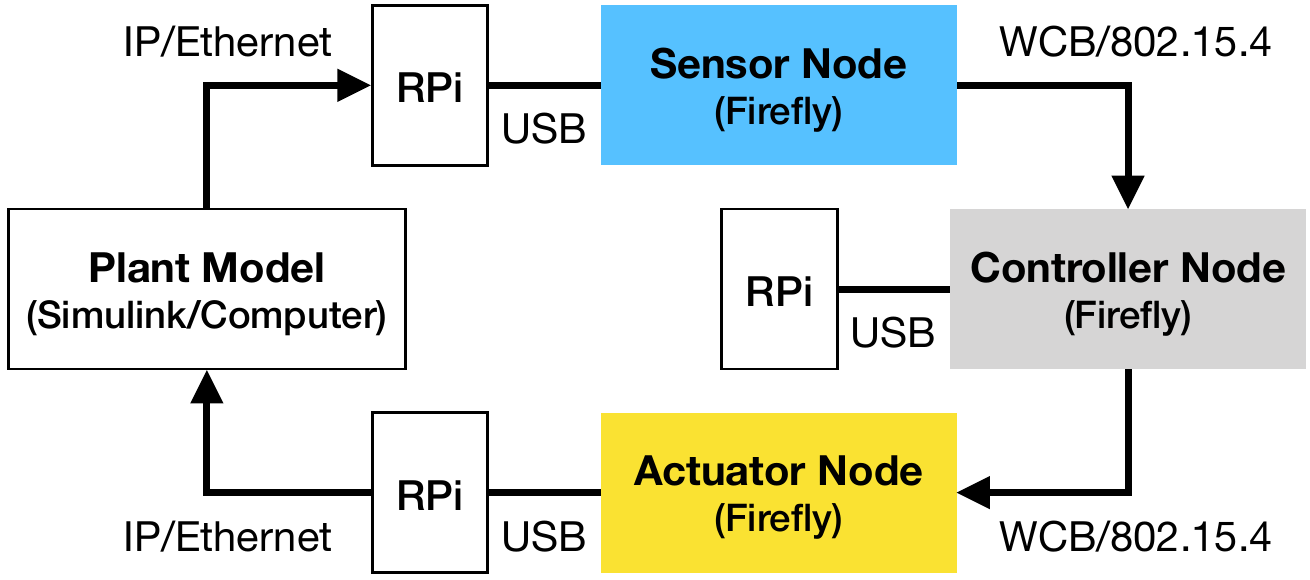}
}\\
  \subfloat[Block diagram of the simulation.] 
  {\label{fig:simulink}
	\includegraphics[scale=0.47]{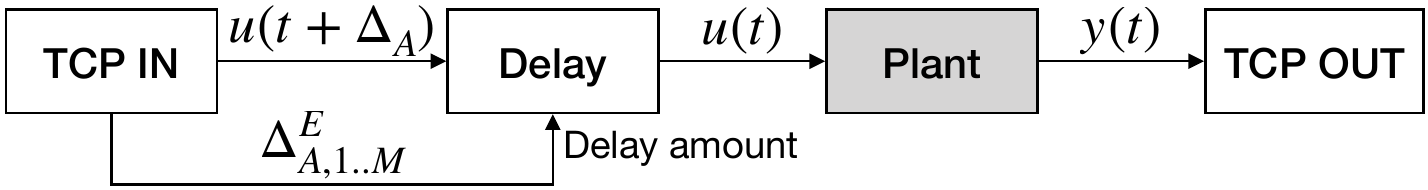}
}
 \vspace*{-2mm}
  \caption{Experimental Framework.}
  \label{fig:dc_model}
 \vspace*{-2mm}  
\end{wrapfigure}

To overcome this limitation, we designed an experimental setup
(Figure~\ref{fig:architecture})
combining a simulated plant with a real
large-scale wireless network. Its architecture is general and can be
applied to systems exploring alternate control strategies and/or
network stacks supporting them.

\fakeparagraph{Real network, simulated plant} The plant model,
implemented in MATLAB/Simulink, emulates the physical system; it
receives actuator state changes as input and produces sensor
readings as output. We replace the real drivers on the wireless
devices with stubs interacting with the plant model, so that
\begin{inparaenum}
\item sensor nodes receive values from the model instead of real
  sensors, and
\item actuator nodes send the commands received from the controller to
  the model instead of the real actuators. 
\end{inparaenum}
Communication between the stubs and the computer running the plant
model occurs out-of-band, via TCP/IP over Ethernet, to avoid
interfering with the wireless network under study.
The latter runs \wcb unmodified, providing multi-hop communication
among sensor, actuator, and controller nodes distributed across large
testbed areas.
Each network node consists of a Zolertia Firefly~\cite{Firefly}, the
actual embedded platform under test, connected via USB to a
Raspberry~Pi (RPi). The Firefly is equipped with a TI~CC2538 SoC
combining an ARM Cortex-M3 MCU and a 2.4GHz IEEE~802.15.4 radio. Our
\wcb prototype is built atop a Contiki~OS port of \glossy for this
SoC~\cite{glossy-cc2538}. The RPi supports the above out-of-band
channel between the Firefly board and the plant model, as well as
enables the automation and remote execution of tests.

\fakeparagraph{Dealing with time} For our setup to provide a realistic
evaluation, it is crucial that the plant simulator, controller, and
wireless network share the same notion of time. The main challenge is
to realign the physical time the last two physical components rely on
with the synthetic one in the plant simulator. Moreover, the
out-of-band Ethernet bridging the real and simulated components is
affected by random delays not present in a real system, which must be
accounted for.

We address these issues as follows. First, we observe that, thanks to
the synchronization inherent in \wcb and other \glossy-based
protocols, all wireless nodes, notably including the controller, share
the same time reference with ms-level accuracy. Therefore, they can
timestamp local events and perform their actions at specified instants
in \emph{global} time. Second, the joint operation of control and
network is \emph{periodic} and \emph{structured}:
\begin{inparaenum}
\item (short) active periods where communication occurs are
  interleaved with (long) periods where the system is quiescent, and 
\item during active periods, the interleaving of communication and
  control follows a well-defined pattern known a priori.
\end{inparaenum}
Third, we leverage the presence of a simulated component to realign
the physical and synthetic time references, precisely by exploiting
the periodic and structured system nature.  During the inactive
portion of the schedule, the simulator runs at its own (faster) pace,
generating the inputs to be fed to physical components at appropriate
(global) times.

\begin{figure}[b!]
  \vspace*{-2mm}  
  \centering
  \includegraphics[scale=0.5]{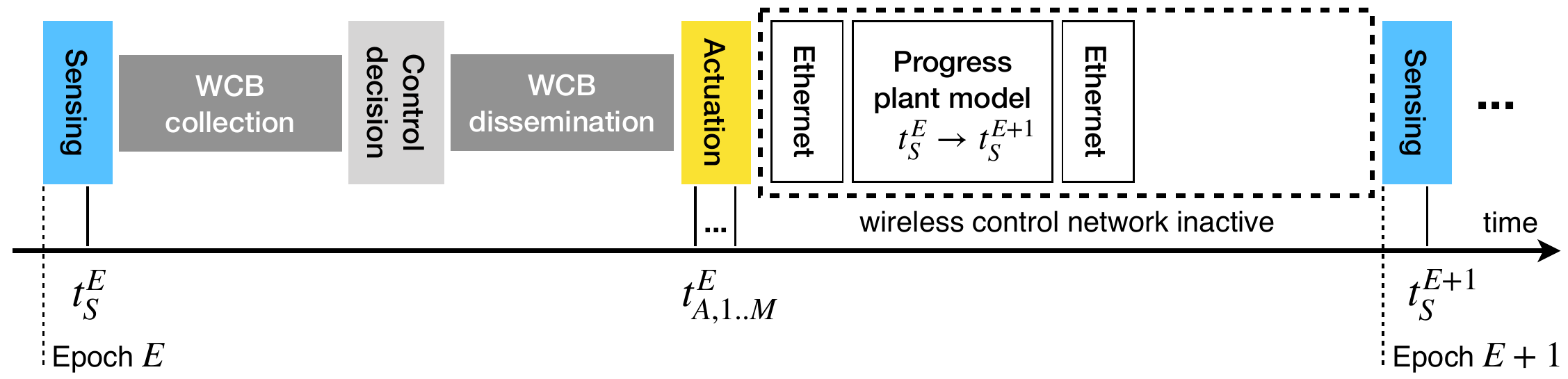}
  \vspace*{-2mm}
  \caption{Synchronous test execution with real wireless network and simulated plant.}\label{fig:synch_eval}
\end{figure}

Figure~\ref{fig:synch_eval} illustrates our strategy. Sensor
acquisition during epoch $E$ occurs at its start time,
$t{_S}^{E}$. The \wcb collection schedule unfolds and, after the
recovery phase, the controller executes and generates the actuation
commands. These are sent during the \wcb dissemination
phase, and received by each actuator $i \in \{1,\ldots,M\}$ at a
potentially different time $t_{A,i}^{E}$. Once dissemination is
complete, the \wcb network enters sleep.
During this inactive period, the actuator stubs send the received
commands to the plant model over the out-of-band network, along with
the reception times $t_{A,i}^{E}$ that, like $t{_S}^{E}$, are
precisely timestamped, as per our first observation. These actuator
states are collected at the computer running the plant model and input
to Simulink, which executes the block diagram shown in
Figure~\ref{fig:simulink} with a simulation time synchronized with the
epoch start, $t{_S}^{E}$.  The timestamps $t_{A,i}^{E}$ are used to
``replay'' the arrival of the actuation commands $u_i$ by taking into
account the \emph{real} delays
$\Delta^E_{A,i} \coloneqq t^E_{A,i} - t^E_S$. Based on this input
vector $u_{i}(t+\Delta^E_{A,i})$, $i \in \{1,...,M\}$, the simulator
advances the model execution in the time interval
$[t_{S}^{E}; t_{S}^{E+1}]$, generating the sensor readings for the
acquisition at the beginning of the next epoch. These are sent to the
stubs on the sensor nodes via the out-of-band network; when the
(physical) time $t_{S}^{E+1}$ arrives, the sensor nodes wake up and
``acquire'' these sensor readings. The process repeats in each epoch.

Nevertheless, the inactive period of the wireless network must
accommodate the worst-case delays induced by model computation and
Ethernet communication. Although we designed our testbed to stop upon
detecting a violation of this requirement, this never happened in our
experiments, where delays ($<$2~s) are significantly smaller than the
control period (60~s). In cases where the control period is shorter
than the delays, execution can be artificially slowed down by
increasing the inactive period and removing the extra \emph{empty}
time in post processing. The opposite, i.e., shortening the inactive
period and adding empty time in post processing, can also be done; we
actually adopted this technique to speed up the execution of our
experiments.

\begin{figure}[!b]
  \vspace*{-4mm}
  \subfloat[\tbdisi] {\label{fig:36node}\includegraphics[scale=0.65, trim=25 0 15 0mm, clip]{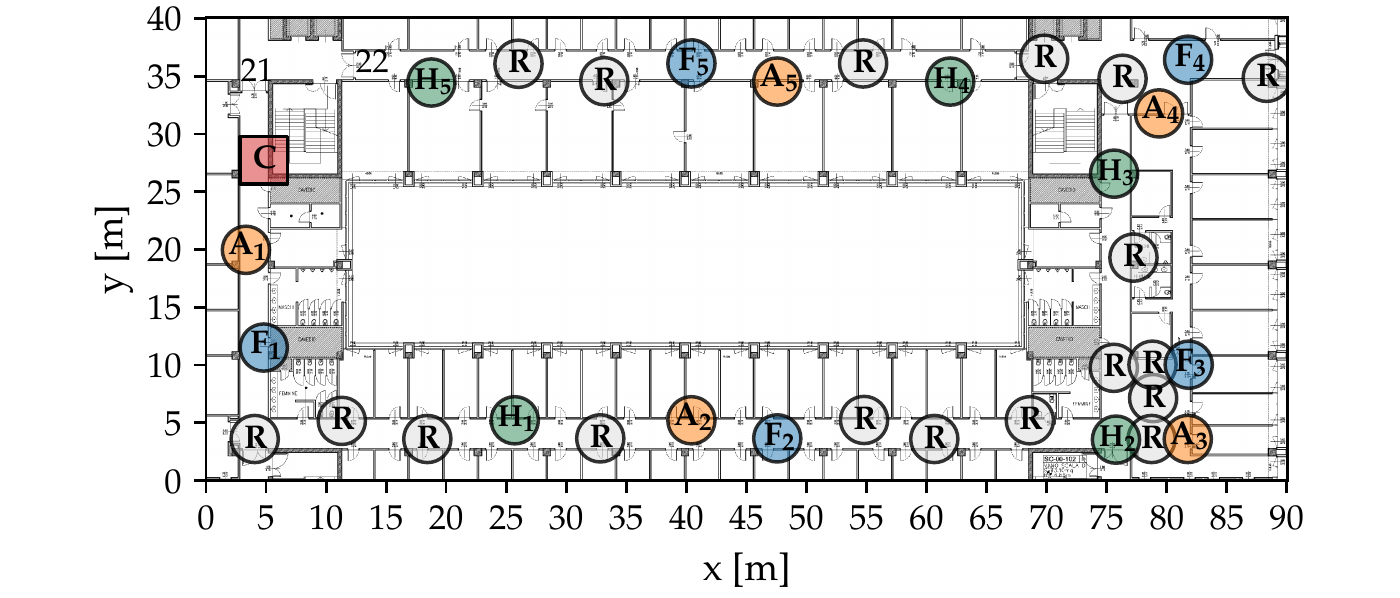}}
  \subfloat[\tbground] {\label{fig:19node}\includegraphics[scale=0.65, trim=15 0 20 0mm, clip]{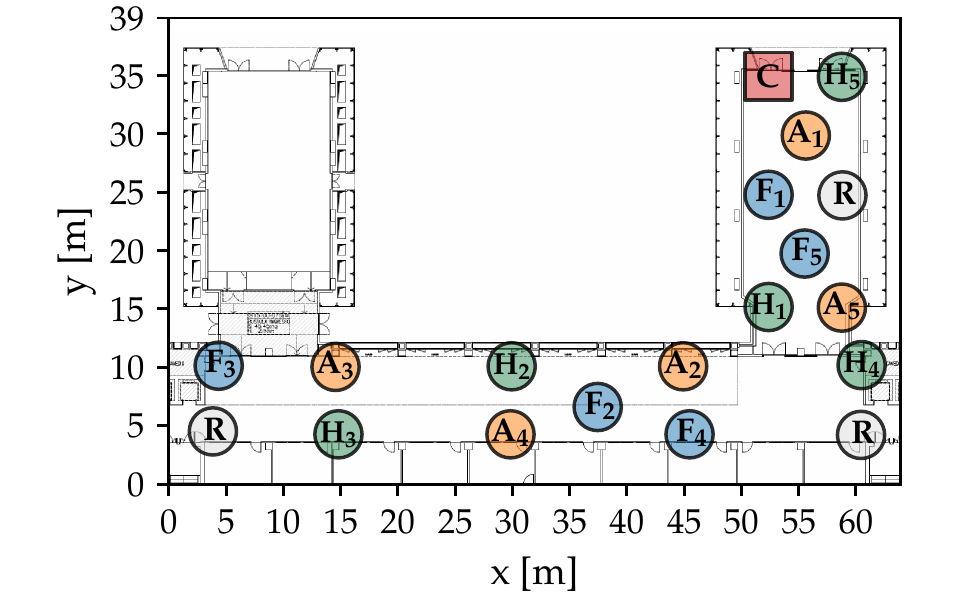}}
  \vspace{-2mm}
  \caption{The wireless testbeds used in our experiments. The red
    square denotes the controller (C), orange circles are actuators
    (A), while light blue and green circles are flow (F) and height
    (H) sensors, respectively. Nodes acting as forwarders (R) are in
    grey. Nodes 21--22 are disabled to increase the network diameter.}
  \label{fig:testbeds}
\end{figure}

\fakeparagraph{Wireless testbeds} We rely on two large-scale multi-hop
wireless testbeds at our premises, called \tbdisi and \tbground,
constituted by 36 and 19 nodes, respectively.
\tbdisi (Figure~\ref{fig:36node}) is deployed along office corridors,
yielding a mostly linear topology spanning a 83$\times$33~m$^2$
area; by disabling \mbox{node~21--22} we enforce a 5-hop network.  \tbground
(Figure~\ref{fig:19node}) is denser and spans a 56$\times$30~m$^2$
L-shaped area; nodes in the same segment are within communication
range, yielding a 2-hop network.

The role of each node (Figure~\ref{fig:testbeds}) mimics our WIS test
case (Figure~\ref{fig:pool_radio}): the actuator and flow sensor nodes
of pool $i$ are close to each another, while the height sensor is far
from them, at the end of pool $i$ and closer to the actuator and flow
sensor of pool $i+1$. Instead, the controller node position maximizes
hop distance, creating a challenging topology for our evaluation.

\fakeparagraph{Benefits and applicability} Our experimental setup is a
contribution offering several advantages. It is \emph{flexible},
enabling experimentation with control systems exhibiting diverse
requirements and time scales by simply developing appropriate Simulink
models. It is easily \emph{replicable} and \emph{scalable} as it does
not require specific hardware components apart from mote-class and
RPi-class devices; existing wireless testbeds~\cite{Flocklab, Indriya,
  competition17} could easily support it. Finally, and most
importantly, it fosters \emph{repeatability}, as the control plant is
simulated, hence not subject to the vagaries of a real system.

        
\section{Configuring (and Improving) the Wireless Control Bus}
\label{sec:protocolconfig}

We empirically study how the parameters of \wcb affect its
performance, and determine the configuration used in the
evaluation. This is also an opportunity to identify low-level
optimizations further improving performance.
Table~\ref{tab:params} summarizes the key parameters, following the
protocol description (\ref{sec:design}). \emph{Slot} parameters govern
the behavior of a single \glossy flood, and can be tuned for each slot
type. \emph{Epoch} parameters govern the use of these slots inside the
active period in each epoch. The table does not consider the number
\Ttb of data collection slots \TF, one per sensor node, as this is an
\emph{application} parameter and therefore only known at deployment time.

\begin{table}
\begin{minipage}[t]{.5\textwidth}
\centering
\caption{Protocol parameters.}\label{tab:params}
\vspace*{-2mm}
\begin{tabular}{c|c}
  \hline
  \multicolumn{2}{c}{\textit{Slot parameters, defined for every slot type}} \\
  \hline
  $W$ & Slot duration \\
  $N$ & Number of packet retransmissions within the slot \\
  \hline
  \multicolumn{2}{c}{\textit{Epoch parameters}} \\
  \hline
  \Ttab & Max. number of \TAF pairs in the recovery phase \\
  \Ctb & Number of command dissemination slots \CTRF \\
    \Etb & Number of event slots \EVF (only \etcpro) \\
\end{tabular}
\end{minipage}
\hspace*{4mm}
\begin{minipage}[t]{.5\textwidth}
\centering
    \setlength\tabcolsep{3.8pt}
    \caption{Reliability of the \wcb configuration.
    }\label{tab:dedicated}
    \vspace*{-2mm}
    \begin{tabular}{c|ccl|ccl}
      \multicolumn{1}{c|}{Slot} & \multicolumn{3}{c|}{ \tbground} & \multicolumn{3}{c}{\tbdisi} \\
      type     & $N$ & $W$ &  \pdrnet & $N$ & $W$ & \pdrnet \\
       \hline
       \SF      & 3 & 7 &  0.99996          & 3 & 10 & 0.99993        \\
       \TF      & 2 & 6 &  0.9994    & 2 & 9  & 0.99914  \\
       \AF      & 3 & 8 &  1.0              & 3 & 11 & 0.99994        \\
       \CTRF    & 2 & 8 &  0.99987          & 2 & 11 & 0.9998         \\
    \end{tabular}
\end{minipage}
\end{table}

\fakeparagraph{Methodology} We determine the parameter values as
inspired by~\cite{crystal}.
We analyze the sensitivity of \wcb to each parameter value via
thousands of floods performed with the same topology, initiating nodes
and packet size as in our evaluation (\ref{sec:eval}). An exception is
the duration $W_x$ of each slot type
$x \in \{\SF, \TF, \AF, \EVF, \CTRF\}$, determined analytically based
on the corresponding number $N_x$ of packet retransmissions and
knowledge of network diameter, packet on-air duration, and \glossy
delay between packet RX and TX, plus a small slack accounting for
potential collisions.

\fakeparagraph{Slot parameters} Table~\ref{tab:dedicated} shows the
configuration we select along with the corresponding mean packet
delivery rate \pdrnet for the \emph{whole} network (opposed to the
sink only). \EVF slots are not reported here as they are used only in
\wcbe; they are analyzed at the end of the section.

A value $N\in\{2,3\}$ ensures very good reliability; higher values
increase consumption without much improvement.
We select $N=3$ for \SF and \AF slots as these are
\begin{inparaenum}[\em i)]
\item crucial to the overall reliability of \wcb, and
\item scheduled \emph{once} per epoch, bearing a moderate
  impact on energy consumption \wrt $N=2$.
\end{inparaenum}
As for \TF slots, they 
\begin{inparaenum}[\em i)]
\item are the largest component of an epoch active portion, always
  present in \perpro and dynamically triggered in \etcpro, and
\item benefit from the safety net of acknowledgements and 
  retransmissions scheduled on-demand during the recovery phase.
\end{inparaenum}
Therefore, we privilege energy consumption over reliability and use
$N=2$. Nevertheless, Table~\ref{tab:dedicated} shows that this value
still achieves a remarkable three-nine reliability of \TF slots
\emph{over the entire network}.

Knowledge of this reliability enables us to estimate analytically the
probability to collect at the sink \emph{all} the $K$ sensor readings,
assuming packet loss modeled as a series of independent and
identically distributed (i.i.d.) Bernoulli
trials~\cite{Glossy-Bernulli}. In our case (\ref{sec:testcase}), this
yields a probability to deliver all $K=10$ sensor readings of 99.3\%
and 99.6\% in \tbground and \tbdisi, respectively.
In other words, at least one reading is lost only in 4--7 epochs out of
1000. In these relatively rare cases, the recovery phase is
automatically triggered, and the lost packets retrieved when needed,
much more efficiently than by increasing the reliability (and consumption)
of \emph{every} \TF flood.

\fakeparagraph{Epoch parameters} In the recovery phase, $\Ttab$ is the
number of \TAF pairs enabling nodes to retransmit packets not
acknowledged by the sink, if any. This parameter directly affects the
reliability of data collection but also the latency of actuation
commands, as their dissemination is always scheduled after the maximum
duration of the recovery phase
(Figure~\ref{img:protocol_struct}). Hereafter, we use $\Ttab=3$ as
we verified experimentally that, in our setup, the probability to lose
$>$3 packets in the collection phase is $<$10$^{-7}$.

On the other hand, the dissemination phase must also be reliable in
addition to timely, as it is crucial to the control operation that
actuation commands are correctly received network-wide. Nevertheless,
a safety net of acknowledgments and retransmissions, akin to the one
supporting many-to-one data collection traffic, would be inefficient
for one-to-many dissemination. Fortunately, a simple and effective
redundancy strategy where the \CTRF slot containing actuation commands
is always repeated $\Ctb$ times is possible. Table~\ref{tab:dedicated}
shows that $N=2$ already makes it unlikely that an actuation message 
is lost network-wide. The probability that the packet is lost 
\emph{multiple} times in a row is therefore very low; we verified 
empirically and analytically that the value $\Ctb=2$ used hereafter
is sufficient to obtain between 6- and 7-nine reliability in our 
testbeds.

\fakeparagraph{Event phase} The reliability of the event phase in
\wcbe is crucial to the correct and timely operation of ETC. 
Nevertheless, the \EVF slots constituting this phase have 
peculiar characteristics. First, they are \emph{shared}; 
several sensor nodes may detect at the same time a violation 
of the triggering condition and decide to signal an event by
concurrently transmitting in the same \EVF slot. Second, their
reception triggers a reaction \emph{at the sink and all sensor nodes},
signaling the need to perform a collection phase. Third, as in the
case of actuation commands, this traffic pattern is not amenable to
acknowledgments, and therefore must rely on alternative reliability
mechanisms.

Table~\ref{tab:eventphase} analyzes the reliability of \EVF slots,
similarly to what reported for the other slots in
Table~\ref{tab:dedicated}, this time considering also a number $U$ of
randomly-selected sensor nodes transmitting in the same shared slot.
Results show that while most of the network, including the sink,
enjoys near-perfect reliability, a few nodes instead experience
repeated losses. This is exacerbated as $U$ increases, with a minimum
network-wide $\pdrnet=97\%$. Unfortunately, losing 3~events out of 100
is unacceptable, as it could hamper ETC performance.

A redundant strategy, similar to the one adopted for the dissemination
phase, mitigates the problem; repeating the \EVF slot for $\Etb=2$
times improves reliability in all configurations and yields a minimum
$\pdrnet=99.3\%$. Increasing $\Etb$ would improve further, but also
severely reduce the energy efficiency of the ETC system, as the event 
phase is scheduled in every epoch of \wcbe.

However, an alternative, energy-efficient technique is possible. We
observe that event packets do not carry data; their mere reception is
what informs nodes that an event has been reported. Consequently,
instead of requiring correct reception of event packets, we consider
the reception of \emph{any} {\ieeestd frame} (even corrupted ones) in
an \EVF slot as an indication of an event detection.

The impact of this technique is beneficial, as shown in the right-hand
side of Table~\ref{tab:eventphase}, reporting the average,
network-wide signal detection rate \sdrnet. Reliability is increased
in all configurations, with a minimum $\sdrnet=99.8\%$ with $U=10$
senders in \tbdisi. Further, reliability rapidly increases as $U$
decreases, achieving or approaching 5~nines. In practice, in our
representative test case the number of sensors concurrently detecting
events is $<$1.2 on average, and always $<$6.

On the other hand, relying on corrupted packets in the \EVF slot may
lead nodes to falsely presume an event has been detected, wasting
energy by incorrectly triggering data collection. We
verified empirically both in our dedicated experiments as well as in
the overall evaluation (\ref{sec:eval}) that the rate of these false
positives is $<$0.003\%, bearing a negligible impact on energy
consumption.

Table~\ref{tab:protocolconf} summarizes the configuration used in the
evaluation.

\begin{table}
\begin{minipage}[t]{.6\textwidth}
\centering
\caption{Reliability of the \EVF phase in \wcbe.}
\label{tab:eventphase}
\vspace*{3mm}
\begin{tabular}{c|ccc|cc|cc}
  \multicolumn{1}{c|}{\multirow{2}{*}{}} &\multicolumn{3}{c|}{} & \multicolumn{2}{c|}{\pdrnet} & \multicolumn{2}{c}{\sdrnet} \\
& $N$ & $W$ & $U$  & \Etb=1 & \Etb=2 & \Etb=1 & \Etb=2   \\
\hline
\multirow{6}{*}{\rotatebox[origin=c]{90}{\begin{tabular}[c]{@{}c@{}}\tbground\end{tabular}}}
& 2 & 4 & 1      & 0.9993 & 0.9999990  & 1.0    & 1.0        \\
& 2 & 4 & 2   & 0.992  & 0.99973    & 0.9986 & 0.99999   \\
& 2 & 4 & 3      & 0.985  & 0.9988     & 0.997  & 0.99994    \\
& 2 & 4 & 5      & 0.973  & 0.995      & 0.991  & 0.9995     \\
& 2 & 4 & 7      & 0.969  & 0.993      & 0.988  & 0.999     \\
& 2 & 4 & 10     & 0.976  & 0.997      & 0.989  & 0.999  \\
\hline
\multirow{6}{*}{\rotatebox[origin=c]{90}{\begin{tabular}[c]{@{}c@{}}\tbdisi\end{tabular}}}
& 2 & 6 & 1   & 0.9988 &  0.999997      & 1.0    & 1.0      \\
& 2 & 6 & 2      & 0.996  &  0.99996       & 0.9994 & 0.999997 \\
& 2 & 6 & 3   & 0.993  &  0.99986    & 0.9988 & 0.999993 \\
& 2 & 6 & 5      & 0.991  &  0.9997        & 0.9984 & 0.99998  \\
& 2 & 6 & 7      & 0.987  &  0.9991        & 0.997  & 0.9998   \\
& 2 & 6 & 10     & 0.97   &  0.995         & 0.989  & 0.998    \\
\end{tabular}
\end{minipage}
\hspace{2mm}
\begin{minipage}[t]{.3\textwidth}
  \caption{\label{tab:protocolconf} \wcb configuration
    in~\ref{sec:eval}.  The values $\Twait_x$ are in ms.}
  \vspace*{-1mm}
\begin{tabular}{c|cc}
  Parameter & \tbground & \tbdisi\\
  \hline
  \NtxS & 3  & 3  \\
  \WS & 7 & 10 \\
  \NtxEV & 2  & 2 \\
  \WEV & 4 & 6 \\
  \NtxT & 2 & 2\\
  \WT & 6 & 9 \\
  \NtxA & 3 & 3 \\
  \WA & 8 & 11 \\
  \NtxCTRL & 2 & 2 \\
  \WCTRL & 8 & 11 \\
  \hline
  \Etb & \multicolumn{2}{c}{2} \\
  \Ttab & \multicolumn{2}{c}{3} \\
  \Ctb & \multicolumn{2}{c}{2} \\
\end{tabular}
\end{minipage}
\end{table}


\section{ETC over WCB: A Testbed Evaluation}
\label{sec:eval}

We now ascertain the ability of \wcb to efficiently support ETC by
fulfilling its peculiar requirements in terms of reliability and
latency, necessary to a correct and efficient control, while retaining
the energy savings enabled by ETC adaptive sampling. To offer a
concrete and complete application of ETC over \wcb, we focus on the
WIS test case and execute in our cyber-physical testbed
(\ref{sec:expsetup}) the control strategy we outlined
(\ref{sec:testcase}) atop the \wcbe variant properly configured
(\ref{sec:protocolconfig}). Each experiment has a duration of one full
day (1440 epochs) of simulated time, repeated multiple times.

We compare against periodic control over \wcbp. Although a comparison
of the latter against the state of the art in networking for periodic
control is outside the scope of this paper, we argue that \wcbp is
more performant than the existing CTX-based solutions we survey
in~\ref{sec:relwork}---themselves outperforming conventional
ones---due to the different design and reliability mechanisms, whose
beneficial impact we show here. In any case, given that \wcbp is
essentially a degenerate case of \wcbe (\ref{sec:perpro}) our choice
compares both control strategies against the same protocol framework,
elucidating the key differences without the bias a completely
different network stack would induce.

\subsection{Control Performance}\label{ssec:controlresults}

Each simulated day starts with $x_{1,i} = 0.05$~m,
$x_{2,i}=x_{3,i}=0$~m for each pool $i$ and no disturbance. Off-take
step disturbances are added at pool~5 as in~\cite{li2011stability}:
0$\rightarrow$16 m\textsuperscript{3}/min at minute 180,
16$\rightarrow$34 m\textsuperscript{3}/min at 450, and
34$\rightarrow$0 m\textsuperscript{3}/min at 600.
As the system has time to settle in between and after disturbances,
we observe it both in steady state and during transient, when
perturbed.

We consider
\begin{inparaenum}
\item an ideal scenario where sensors yield perfect readings, and
\item one where independent normally-distributed pseudo-random white
  noise is added to both level and flow measurements, with zero
  mean and standard deviation of 0.001~m and
  1~m\textsuperscript{3}/min, respectively.
\end{inparaenum}
In the ideal scenario, the only source of randomness is the network,
allowing us to isolate the impact of the protocol stack on control
performance. In the second scenario, the added noise introduces
variability (and degradation) of the ETC sampling performance,
enabling a more realistic assessment.

\fakeparagraph{Metrics} We focus on the number of samples generated as
well as on two metrics based on the \emph{integral average error}
(IAE) of a signal $x(t)$ \wrt its reference $x^*$
\begin{equation}\label{eq:IAE}
\mathrm{IAE}(x,x^*,\rev{T_\mathit{exp}}) \coloneqq \frac{1}{\rev{T_\mathit{exp}}}\int_0^{\rev{T_\mathit{exp}}}|x(t) - x^*|\d t.
\end{equation}
This standard control performance metric measures the accumulated
tracking error; the smaller its value, the faster states converge to
their references. In our case $\rev{T_\mathit{exp}}=1440$ minutes, the duration of the
experiments. Since height references are already accounted for in the
variables $x_{1,i}$, we set $x^*=0$, yielding the metrics
$\mathrm{IAE}_i \coloneqq \mathrm{IAE}(x_{1,i},0,\rev{T_\mathit{exp}}).$
For each simulation, we compute the sums and maxima of IAEs over the pools, 
with the following shortened notations:
\begin{equation}\label{eq:metrics}
\mathrm{IAE}_{\sum} \coloneqq \sum_{i=1}^5\mathrm{IAE}_i, \quad
\mathrm{IAE}_{\max} \coloneqq \max_{i\in\{1,...,5\}}\mathrm{IAE}_i,
\end{equation}

\begin{figure}[!t]
	\begin{center}
		\includegraphics{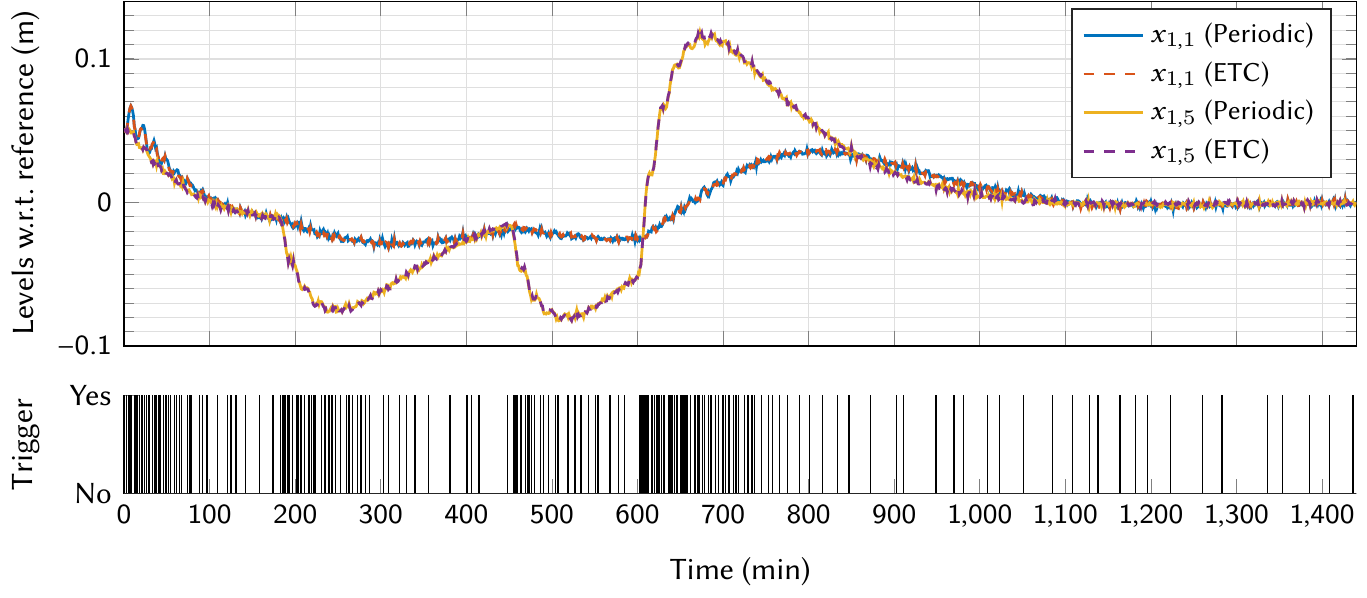}
		\vspace{-2em}
		\caption{\label{fig:control_output} ETC \vs periodic
                  control, both over \wcb in \tbdisi. Top: Level \wrt
                  reference for the 1$^{st}$ and 5$^{th}$ pools, 1-day
                  executions with measurement noise.  Bottom: Sampling
                  instants for the ETC case.}
	\end{center}
\end{figure}

\fakeparagraph{Results} The pool heights follow a similar trajectory
under both control strategies (Figure~\ref{fig:control_output}, top)
and with a similar performance in reference tracking
(Table~\ref{tab:controlresults}), confirming the desirable property
that ETC yields essentially the same control output of periodic
control.  However, \emph{ETC generates significantly fewer samples
  than periodic control}, almost 90\% less in the ideal scenario and
only slightly more, 87\% less on average, with measurement noise
(Table~\ref{tab:controlresults}).  The sample pattern for ETC
(Figure~\ref{fig:control_output}, bottom) highlights that, as
expected, sampling is more frequent when transients are stronger, and
becomes sporadic as the system approaches steady state. 

\newcommand{\ten}[1]{10$^{#1}$}
\newcommand{\timesten}[1]{$\times$10$^{#1}$}
\begin{table}
  \caption{\label{tab:controlresults} Sampling and control performance
    metrics from experiments: mean (standard deviation when different 
    from 0) over 8~executions of 1~day of plant operations each.}

  {\renewcommand{\arraystretch}{1.1}
    \begin{tabular}{cccccc}
      \hline 
      Scenario & Testbed & Sampling & Sample count & $\mathrm{IAE}_{\sum}$ (m) & $\mathrm{IAE}_{\max}$ (m) \\ 
      \hline 
      \multirow{4}{*}{Without noise} & 
      \multirow{2}{*}{\tbground} & ETC & 149 & 0.1084 & 0.03283 \\
      & & Periodic & 1440 & 0.1085~($<$\ten{-6}) & 0.03293~($<$\ten{-6}) \\
      \cline{2-6} 
               & \multirow{2}{*}{\tbdisi} & ETC & 148 & 0.1088~($<$\ten{-6}) & 0.03286 \\
      & & Periodic & 1440 & 0.1085~($<$\ten{-6}) & 0.03293~($<$\ten{-6}) \\
      \hline
      \multirow{4}{*}{With noise} &  
      \multirow{2}{*}{\tbground} & ETC & 186.1~(5.743) & 0.1091~(1.21\timesten{-4}) & 0.03311~(6.1\timesten{-5}) \\
      & & Periodic & 1440 & 0.1088~(3.8\timesten{-5}) & 0.033~(2.1\timesten{-5}) \\
      \cline{2-6}
      & \multirow{2}{*}{\tbdisi} & ETC & 185.4~(4.984) & 0.109~(1.39\timesten{-4}) & 0.03308~(4.7\timesten{-5}) \\
      & & Periodic & 1440 & 0.1088~(3.8\timesten{-5}) & 0.033~(2.1\timesten{-5}) \\
      \hline
    \end{tabular}
  }
\end{table}

\rev{It is important to remark that the savings ETC can provide \wrt
  periodic control are highly dependent on the control problem at
  hand, as the average PETC sampling frequency depends in non-trivial
  ways on the system dynamics, control design, and triggering
  mechanism. The formal computation of this value has only recently
  been made possible~\cite{gleizer2021hscc}. Nevertheless, we} observe
that the \emph{variation} of sample count across experiments, captured
by the standard deviation (Table~\ref{tab:controlresults}), appears in
ETC only in the scenario with measurement noise and is completely
absent in the ideal one. This is a witness of the consistent
performance of \wcbe in terms of reliability and latency, analyzed
next: \emph{practical control aspects like measurement noise induce
significantly higher variations in ETC sampling than the vagaries of
the wireless communication}. \rev{Notably, this enables a desirable
separation of concerns during system development, as the assessment
of the benefits ETC provides, and therefore the decision on whether
or not to employ it for the specific case at hand, can be performed
accurately and entirely during the control design phase.}

\subsection{Network Performance}
\label{sec:eval:nw}

\begin{table}[!t]
  \caption{Performance of \wcb in hardware-in-the-loop testbed
    experiments: mean (and standard deviation when non-zero) over 16
    executions of 1440 epochs each, \ie 1 day of plant operation.}
  \label{tab:protocolresults_noiseless} 
	\begin{tabular}{ccccccc}
		\hline
		\multicolumn{1}{c}{Testbed} & \begin{tabular}[c]{@{}c@{}}Protocol \end{tabular} & \begin{tabular}[c]{@{}c@{}}Event detection\\reliability~[\%]\end{tabular} & \begin{tabular}[c]{@{}c@{}}Data collection\\reliability~[\%]\end{tabular} & \begin{tabular}[c]{@{}c@{}}Actuation\\reliability~[\%]\end{tabular} & \begin{tabular}[c]{@{}c@{}}Latency of actuation\\commands~[ms]\end{tabular} \\ 
		\hline
		\multirow{2}{*}{\tbground} & \multicolumn{1}{l}{\etcpro}& 100 & 100 & 100 & 192.021~(0.04) \\
		& \multicolumn{1}{l}{\perpro} & ---& 100 & 100 & 180.023~(0.02) \\
		\hline
		\multirow{2}{*}{\tbdisi} & \multicolumn{1}{l}{\etcpro} & 100 & 100 & 100 & 253 \\
		& \multicolumn{1}{l}{\perpro} & --- & 100 & 100 & 237.017~(0.012) \\
		\hline
	\end{tabular}
\end{table}

The reliability of event detection, sensor reading collection and 
command dissemination, together with the actuation latency, are 
crucial to the control performance we observed.

Table~\ref{tab:protocolresults_noiseless} reports the average of these
metrics across 16~test runs, i.e., 1440$\times$16=23040 epochs in each
row. \wcb achieves zero packet losses regardless of the functionality,
protocol variant, and testbed considered, confirming the effectiveness
of its strategy (\ref{sec:design}) and configuration
(\ref{sec:protocolconfig}).
Recovery mechanisms are key to achieve this result. Log inspection
shows that, for data collection, they are triggered $\sim$1\% of the
times; while small in absolute terms, this fraction of lost packets,
if not recovered, \rev{would make ETC trigger more often than needed
  and potentially degrade performance.}  

\rev{These losses, mainly caused by interference from WiFi access
  points and devices in the indoor office spaces where our testbeds
  are deployed, are effectively and efficiently mitigated by our use of
  channel hopping. Our technique is directly inspired by work on
  Crystal~\cite{crystal2} where it has been shown capable to
  withstand significantly stronger interference. Further, \emph{all}
  of the top-three winning systems (including Crystal) in the 2018
  and 2019 editions of the EWSN Dependability Competition rely on
  some form of channel hopping to overcome its nearly unreasonable
  noise levels. Therefore, while we cannot offer an evaluation of
  \wcb under strong interference as in~\cite{crystal2}, prohibitive
  both in terms of testbed logistics and text limitations, we
  incorporate in our system the state-of-the-art techniques for
  interference resilience, an aspect entirely neglected by 
  existing network stacks for
  ETC~\cite{Manuel_AWCS,kartakisfu2018commschemes,van2016experimental,dolk2017platoon}
  whose other shortcomings we discuss in~\ref{sec:relwork}. 
  Moreover, although the sparser and aperiodic traffic induced by
  ETC increases the importance of each packet, this control strategy
  is intrinsically resilient to packet loss. When this occurs, the
  effect is simply the triggering of more events due to incomplete
  information at the controller; this transiently impacts energy
  consumption but not the correctness of control, as long as the
  required sensor readings are delivered at the controller within a
  maximum tolerable delay.}

  The latency between the beginning of an epoch and the delivery
  of the \emph{last} actuation command is also very small,
  especially if compared to the sampling period (hundreds of ms
  \vs 60~s). Further, it has minimal jitter, as commands usually
  reach actuators in the first \CTRF slot. Interestingly, the
  different network diameter of the two testbeds induces an
  inevitable difference in the latency of actuation
  commands. 
  Although this difference is very small ($<$61~ms) \wrt the
  system dynamics (hours), the ETC sampling patterns are known
  to be sensitive to small perturbations over the long run;
  however, the net effect is only a small difference in the ETC
  sample count (Table~\ref{tab:controlresults}).

Finally, as expected, \etcpro is slightly slower ($\sim$6.7\%) than
\perpro due to the additional event detection phase, although the
absolute difference is negligible \wrt the sampling period and does
not affect the control output, as already mentioned
(Figure~\ref{fig:control_output}, Table~\ref{tab:controlresults}).

\subsection{Energy Consumption}

The wireless transceiver is notoriously the most power-hungry
component in networked embedded systems, and the one whose
contribution ETC seeks to minimize. Therefore, we compare ETC \vs
periodic control in terms of the radio duty-cycle
$\dc=\frac{\Ton}{\rev{T_\mathit{exp}}}$, i.e., the per-node radio-on
  time over experiment duration, a metric commonly accepted as a
  reliable proxy for energy consumption.

\fakeparagraph{Key finding} Table~\ref{tab:energy_comp} confirms that
our embodiment of ETC consumes significantly less than periodic
control---one of our goals. The reason lies precisely in the
\emph{interplay} between ETC and the network stack supporting its
operation, \wcbe. By design, ETC abates traffic by triggering sensor
data transmissions only when needed for control. In our test case,
$>$89\% of the periodic samples are suppressed in the ideal case, and
$>$87\% in the noisy one. In general, this traffic suppression does
not automatically translate in energy savings. Nevertheless,
\etcpro minimizes consumption when the system is in steady state while
ensuring timely and reliable communication when required to support
control. In our case, this yields a \dc reduction $>$62\%, with
marginal differences in the two testbeds due to their different
network diameter. 
Therefore, \emph{\etcpro effectively translates the significant
  reduction of control traffic achieved by ETC into corresponding
  savings in energy consumption}. This is a significant leap forward
\wrt state-of-the-art ETC
literature~\cite{wang2008event,girard2014dynamic,heemels2013periodic,mazo2011decentralized,tabuada2007event}
whose energy reduction is hampered by inefficient protocols and
limited to small-scale star topologies.

\begin{table}[!t]
\centering
\caption{\label{tab:energy_comp} Sampling and duty-cycle performance
  of ETC and periodic control \vs presence of measurement
  noise. Results are average percentages over 8~executions of 1440
  epochs each, i.e., 1 day of plant operation.}
\label{tab:DCtab}
\vspace*{-2mm}
\begin{tabular}{c|c|c|c|c|c|c|c}
  \multicolumn{1}{}{}  &  \multicolumn{1}{|}{} & \multicolumn{3}{|c|}{\textbf{No measurement noise}} & \multicolumn{3}{c}{\textbf{With measurement noise}} \\
  \hline
  \multirow{2}{*}{Testbed}  & \multirow{2}{*}{Control} & \multirow{2}{*}{\dc} &  \multicolumn{2}{c|}{reduction} & \multirow{2}{*}{\dc} & \multicolumn{2}{c}{reduction} \\
    & & &  sampling & \dc & & sampling  & \dc  \\
  
	\hline
	\multirow{2}{*}{\tbground} & ETC & 0.0319 & \multirow{2}{*}{89.65} & \multirow{2}{*}{67.84} & 0.0341 & \multirow{2}{*}{87.02} & \multirow{2}{*}{65.45} \\
    & Periodic & 0.0992 & & & 0.0987 & & \\
\hline
	\multirow{2}{*}{\tbdisi} & ETC & 0.0413 & \multirow{2}{*}{89.72} & \multirow{2}{*}{64.58} & 0.0438 & \multirow{2}{*}{87.13} & \multirow{2}{*}{62.47} \\
	& Periodic & 0.1166 & & & 0.1167 & & \\
\hline
\end{tabular}
\end{table}

\begin{table}[]
  \caption{\label{tab:radio-on} Average per-epoch radio-on time \Ton
    and duty-cycle \dc without measurement noise. Values are the
    average over 8~executions of 1440~epochs each, i.e., 1 day of plant
    operation.}
  \vspace*{-2mm}
	\begin{tabular}{cc|ccccc|c}
      \multicolumn{1}{c}{} & \multicolumn{1}{c}{} & \multicolumn{5}{|c}{\textbf{\etcpro}}  & \multicolumn{1}{|c}{\textbf{\perpro}} \\
 		\hline
                Testbed & Metric & \begin{tabular}[c]{@{}c@{}}No event\\ detected\end{tabular} &
		\begin{tabular}[c]{@{}c@{}}Event\\detected\end{tabular} &
		\begin{tabular}[c]{@{}c@{}}Transient\\(600--750)\end{tabular} &
		\begin{tabular}[c]{@{}c@{}}Steady state\\ (1000--1440)\end{tabular} &
		\begin{tabular}[c]{@{}c@{}}1 day\\(0--1440)\end{tabular} &
		\begin{tabular}[c]{@{}c@{}}1 day\\(0--1440)\end{tabular} \\ 
		\hline
                \multirow{2}{*}{\tbground} & \Ton~[ms] & 13.81 & 65.51 & 29.58 & 14.40 & 19.16 & 59.50 \\
                        & \dc~[\%]       & 0.0230 & 0.1092 & 0.0493 & 0.0240 & 0.0319 & 0.0992 \\
                \hline
                \multirow{2}{*}{\tbdisi} & \Ton~[ms] & 18.82 & 76.93 & 36.58 & 19.60 & 24.79 & 69.98 \\
                        & \dc~[\%]       & 0.0314 & 0.1282 & 0.0610 & 0.0327 & 0.0413 & 0.1166\\
		\hline
	\end{tabular}
\end{table}

\begin{figure}[!b]
  \centering
  \includegraphics[width=\textwidth]{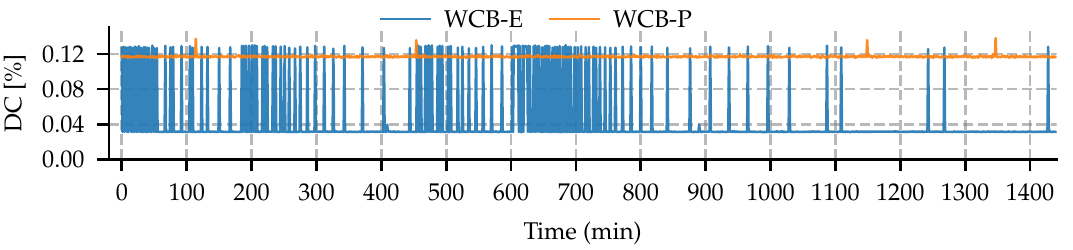}
  \vspace*{-8mm}
  \caption{Comparison of the average network duty-cycle per-epoch of \etcpro
    and \perpro during one day of plant operations in \tbdisi in absence
    of measurement noise.
  }
  \label{fig:DC_comparison}
\end{figure}

\fakeparagraph{Dissecting the energy contribution}
Figure~\ref{fig:DC_comparison} highlights where energy savings arise
from, by comparing the average \dc per epoch of \perpro and \etcpro 
across one day of plant operation.
The behavior of the periodic controller is invariant \wrt
system conditions. Therefore, \perpro must acquire sensors readings
and disseminate actuation commands in \emph{every} epoch, resulting in
a nearly-constant \dc; the small spikes correspond to occasional
recovery phases. In contrast, the adaptive ETC controller triggers
communication via \wcbe only when needed. This results in a pattern
similar to Figure~\ref{fig:control_output}, although here we focus on the
ideal case as it simplifies observations concerned with communication
by separating them from measurement noise.  After the initial settling
phase, Figure~\ref{fig:DC_comparison} clearly shows how \dc increases
in conjunction with off-take step disturbances (minute 180, 450, and
600) and reduces when the system approaches stability (1000--1440).

Table~\ref{tab:radio-on} offers additional insights on \Ton and \dc,
by comparing the invariant control operation of \wcbp against the
various stages of ETC operation over \wcbe.
In epochs where no event is detected \etcpro saves 73.1\% and 76.8\%
\wrt \perpro in \tbdisi and \tbground, respectively. Energy is
minimized by putting the network to sleep right after the \EVF phase
(Figure~\ref{img:protocol_struct}). Otherwise, when an event is
detected \etcpro is slightly more active ($\leq 10.1\%$) due to the
extra \EVF slots.

\fakeparagraph{Generalizing to other scenarios} These results show how
the efficiency of ETC over \wcbe ultimately depends on how frequently
the triggering condition is violated. As long as events are relatively
\emph{rare}, the energy savings in steady-state outweigh the extra
cost of the \EVF phase.

System designers must ascertain this tradeoff in the early stages of
development, to select the most appropriate control strategy and the
corresponding network stack supporting it.
Luckily, analytical models for the energy consumption of both \wcb
variants can be easily derived, as all nodes follow the same, global,
periodic schedule.
Once the average network-wide radio-on time \tfX of each slot type is
estimated as in~\cite{crystal} and~\ref{sec:protocolconfig}, the
overall per-epoch radio-on time \Tfpp of \perpro is simply the sum of
\tfX across slots in each protocol phase, invariant \wrt event
detection. The one for \etcpro is then derived as:
\begin{equation*}\label{eq:ton}
\Tfep = \pev \times (\Tfpp + \Etb \times \tfEV) + (1-\pev) \times (\tfS + \Etb\times \tfEV)
\end{equation*}
where \pev is the average frequency of epochs with at least one event
and \Etb the number of \EVF slots (\ref{sec:protocolconfig}).  \dc is
computed for both cases by dividing the radio-on time by the epoch
duration \rev{\epoch}.

Figure~\ref{fig:dc_model} exemplifies the tradeoffs at stake by
reusing the parameters from our evaluation except for the frequency
\pev, whose value here is varied to represent, in an abstract setting,
the \dc resulting from several hypothetical control problems. The
charts show how, in these conditions, periodic control over \wcbp
becomes preferable \vs ETC over \wcbe only when $\pev \geq90\%$; the
latter enables energy savings even when \pev approaches this
break-even point. For instance, when $\pev \approx 70\%$, \dc is
reduced by nearly 15\%, which becomes 25\% when $\pev \approx 60\%$,
still extending system lifetime significantly.  Overall, this confirms that
ETC over \wcbe supports a wide range of real-world control problems
and systems where it unlocks remarkable energy savings, ultimately
pushing the envelope of the application of cyber-physical systems to
untethered scenarios.

\begin{figure}
  \subfloat[\tbdisi] {\label{fig:dc_model_disi}\includegraphics[width=0.5\columnwidth]{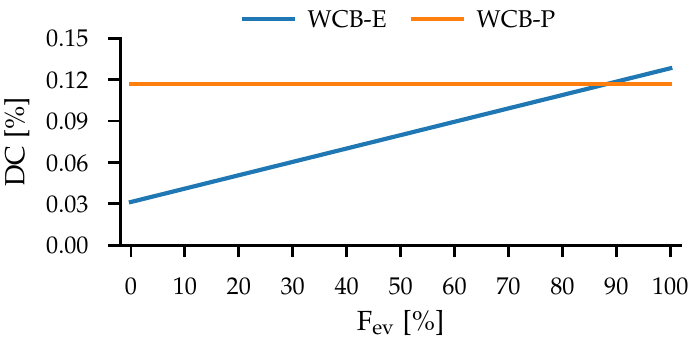}}
    \subfloat[\tbground] {\label{fig:dc_model_ground}\includegraphics[width=0.5\columnwidth]{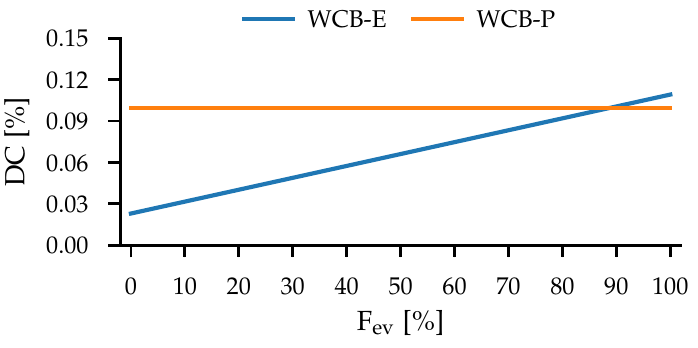}}
  \vspace*{-2mm}
  \caption{Comparison of \perpro and \etcpro \vs the frequency of epochs
    with events, in both testbeds.}
  \label{fig:dc_model}
\end{figure}

\rev{\fakeparagraph{Implications of epoch duration selection} The
  value of \epoch is a crucial parameter that determines a trade-off
  between control responsiveness and energy consumption.
  From a control design standpoint, \epoch
  should be as small as possible to achieve the best control
  performance; however, from a communication standpoint, this causes a
  corresponding increase in duty cycle for both ETC and periodic
  control. Another aspect to be considered is that a small \epoch
  typically leads to fewer events generated in the epoch, i.e., a
  smaller \pev, increasing the relative benefit of ETC \wrt periodic
  sampling (Figure~\ref{fig:dc_model}). Table~\ref{tab:epochvsdc}
  offers a concrete example of these tradeoffs by showing how duty
  cycle changes in our scenario with an epoch duration smaller than
  the value $\epoch=60$~s used here. The values of \dc for ETC and
  periodic control are estimated from simulations using the model for
  \tbdisi in Figure~\ref{fig:dc_model}. As shown in the table, when
  aiming to minimize energy consumption, a general guideline would be
  to set \epoch to the highest value ensuring that control performance
  is within specifications. This is the criterion we adopted here,
  setting the value $\epoch=60$~s to match the fundamental sampling
  period $h$ recommended in the literature~\cite{weyer2001system}
  (\ref{sec:wis:control}), which guarantees a control performance
  within 1\% of the nominal continuous-time one. }
	
	\begin{table}
          \begin{center}
			\caption{\label{tab:epochvsdc} \rev{Effects of the epoch duration on duty cycle, for ETC and periodic control. The ETC savings in \dc relative to periodic are computed as $1 - \frac{\dcetc}{\dcperiodic}$.}}
			\vspace*{-3mm} \rev{
			\begin{tabular}{c|cccccc}
				\hline
				\epoch (s) & \# Events & \# Epochs & \pev (\%) & \dcetc (\%) & \dcperiodic (\%) & ETC savings in \dc (\%)\\
				\hline
				60 	&  187 & 1440 & 13 &  0.034 & 0.099 & 65.7 \\
				45  &  195 & 1920 & 10.1 & 0.042 & 0.132 & 68.2 \\
				30  &  211 & 2880 & 7.3 & 0.059 & 0.198 & 70.2 \\
				15  &  234 & 5760 & 4 & 0.106 & 0.397 & 73.3 \\
				5   &  237 & 17280 & 1.4 & 0.290 & 1.190 & 75.6 \\
				1   &  268 & 86400 & 0.3 & 1.397 & 5.950 & 76.5 \\
				\hline
			\end{tabular}
                        }
          \end{center}
    \end{table}

     
\section{Related Work}
\label{sec:relwork}

The adaptive control strategy of ETC raised a lot of interest in the
last decade, with several researchers tackling the design of new
triggering conditions and other strategies to reduce communication
further~\cite{wang2008event,girard2014dynamic}, improve applicability
on digital platforms~\cite{heemels2013periodic}, and decentralize
triggering conditions~\cite{mazo2011decentralized}.
An overview of the state of the art in ETC can be found
in~\cite{heemels12:introduction,miskowicz18:event}.

However, the benefits unleashed in theory by ETC must be confirmed in
practice by real-world testbeds. This is true in
general~\cite{lu16:real} and even more poignant for ETC, given the
peculiar challenges it poses to communication
(\ref{sec:intro},~\cite{marco-smart_manuf}). Unfortunately, only
few works investigate ETC performance via prototypes. These use
\ieeestd~\cite{Manuel_AWCS,kartakisfu2018commschemes},
WiFi~\cite{van2016experimental}, or G5
(IEEE~801.11.p)~\cite{dolk2017platoon}, but always in a single-hop
topology with at most 5~nodes, hardly representative of staple
real-world use cases for WNCS.

In contrast, the work described here is validated with a realistic
setup that combines a model of the system under control with a real,
multi-hop low-power wireless network, yielding a significant level of 
realism to the evaluation. These testbeds are unfortunately rare in
the literature. The closest is the one proposed in~\cite{ICII18},
featuring a similar combination of modeled system and real
network. Nevertheless, the concise description does not detail if and
how network-induced random delays are mitigated; further, it relies
on the PTP protocol for time synchronization, requiring dedicated,
expensive hardware. In contrast, our testbed explicitly targets random
delays with an architecture (\ref{sec:expsetup}) that, in addition,
provides the extra flexibility to speed up or slow down the real-time
execution.
Moreover, it uses commonplace devices and is therefore
easily replicable by other researchers.

Apart from providing a realistic evaluation, in this paper we have
tackled the crux of the matter by proposing a network stack expressly
targeting \rev{the peculiar traffic patterns and requirements induced
by ETC. For these, the stacks} commonly used in industrial control,
e.g., WirelessHART~\cite{wirelessHART}, ISA100.11.a~\cite{ISA},
6TiSCH~\cite{6TiSCH}, do not offer the necessary guarantees \rev{in
terms of timeliness, reliability, and energy-efficiency,
especially in multi-hop configurations. Research proposals exist
that cater for dynamically changing traffic demands, crucial in
ETC; however, they are based on expensive centralized or
distributed scheduling (e.g., as in
DistributedHART~\cite{distHART}) or unrealistic assumptions about
link reliability (e.g., as in D$^2$-PaS~\cite{d2pas}). In
contrast, \ctx-based stacks do not require explicit routing,
per-link scheduling, or continuous link monitoring. This enables
excellent performance along the three performance dimensions above
but has also been exploited to adapt to dynamically changing
traffic demands~\cite{crystal,trobinger20:one}. This state of the
art directly motivates our work and specifically the use of
\ctx.}

In this respect, the design of \wcb is inspired by two systems: the Low-power
Wireless Bus (LWB)~\cite{lwb} and \crystal~\cite{crystal}.
The former was the first to make explicit the
potential of \ctx for abstracting communication into a network-wide
bus, generating several follow-up variants. For instance,
Blink~\cite{Blink} targets hard real-time communications by equipping
LWB with a real-time scheduler based on earliest deadline first. eLWB
extends LWB with the ability to handle events, as a side contribution
of a more general architecture targeting an acoustic emission
monitoring system~\cite{eLWB}. In eLWB, the reaction to the event is
centralized at the controller, while in \wcb it is decentralized at
sensor nodes, yielding lower latency. Further, eLWB focuses on
monitoring rather than control, without dedicated reliability
mechanisms, crucial in ETC and discussed later.

LWB has been exploited also specifically for control. The system
in~\cite{MarcoTCPS} supports feedback control, stability guarantees,
and mode changes over multi-hop wireless networks for systems with
fast dynamics (tens of ms). Latency is therefore the main focus rather
than reliability, for which dedicated mechanisms are not provided. The
paper exploits a periodic controller. Another work by the same group
explores instead self-triggered control~\cite{MarcoSelf-triggered}
where, contrary to ETC, nodes \emph{predict} when they expect to
trigger an event; this information is exploited to reserve the
required communication slots with LWB. Self-triggered control is also
studied in~\cite{ICII18}, and compared against rate adaptation; in
both control strategies, the necessary communication is provided by a
variant of LWB.

The aperiodic, unpredictable communication patterns of ETC are
significantly more challenging than the pre-defined or predictable
ones induced by periodic and self-triggered control. ETC in principle
enables minimal network overhead during quiescent, steady-state
periods, but also demands both timely and reliable communication
otherwise, to guarantee correctness and performance.  \rev{LWB does
not cater for aperiodic communication, let apart guaranteeing its
conflicting requirements \wrt timeliness, reliability, and
energy-efficiency, as we do instead in \wcb. Therefore, none of the
stacks above, directly built atop LWB support these requirements
either; further, none of them provides dedicated mechanisms
\emph{expressly} targeting reliability, as in our case.}

Instead, these conflicting requirements have been reconciled in
\crystal~\cite{crystal,crystal2}. Aperiodic communication ``makes
each packet count'', as it is transmitted unpredictably and
sporadically, implicitly carrying more information. \crystal focuses
on data collection and exploits the capture effect to support
concurrent, reliable transmission of sensor readings, individually
acknowledged by a \glossy flood. This pattern directly inspires the
\TF and \AF slots in \wcb, where they are combined differently. In
\crystal, concurrent senders are a priori unknown; in the worst case
where all $U$ nodes transmit, at least $2U$ \glossy floods are
required. In \wcb, data collection occurs only if and when an event 
signaling a violation of the ETC triggering condition is disseminated. 
As this occurs reliably and in a distributed fashion, it eliminates 
contention and triggers collection, always from \emph{all} sensor nodes, 
using only $U+1$ floods. The recovery phase, reminiscent of the \TAF pairs of
\crystal, must therefore retrieve only an occasional missed packet,
rather than all competing ones, limiting overhead and bounding the
recovery duration, crucial for predictable control operation.


\section{Conclusions and Future Work}
\label{sec:ending}

We presented the Wireless Control Bus (\wcb), the first network stack
efficiently supporting the peculiar communication requirements induced
by ETC. Unlike the few prototypes reported in the literature, \wcb
expressly targets multi-hop, low-power wireless networks, and advances
the state of the art by significantly reducing the gap between
communication savings and energy savings---a well-known issue
hampering ETC adoption.  We design a centralized state feedback
controller using a novel, modified decentralized periodic ETC suited
for step disturbance rejection, combine it with \wcb, and evaluate its
performance in network-in-the-loop setups emulating a 15-state water
irrigation system over a real-world multi-hop network. Our results
show that \wrt periodic control, also implemented over \wcb:
\begin{inparaenum}
\item ETC reduces samples by $>$87\%, translated by \wcb into energy
  savings $>$62\%, and 
\item control performance is essentially equivalent in the two
  strategies and consistent across experiments, witnessing the extreme
  dependability of the network layer.
\end{inparaenum}

We intend to release publicly \wcb as open source. We believe that the
availability and performance of \wcb, unlocking the full potential of ETC,
may fuel new research on this topic. Our own agenda includes exploring the
combination of \wcb with other decentralized ETC
frameworks~\cite{heemels2013periodic}, implementing
theta-adaptation~\cite{mazo2011decentralized}, \rev{and using traffic
models~\cite{kolarijani2016formal} to further reduce energy
consumption by scheduling longer periods of sensor node sleep, along
the lines of~\cite{gleizer2020scalable}.}
Concerning our test case of
water irrigation systems, we are working on alternate control
architectures, like the robust output-feedback controllers
in~\cite{li2011stability}, and developing a testbed using a
scaled-down irrigation channel, to investigate other practical aspects
of wireless ETC. Finally, the exploitation of CTX on radios other than
\ieeestd opens intriguing opportunities. For instance, an ultra-fast
data collection layer has recently been proposed for ultra-wideband
(UWB) radios~\cite{trobinger20:one}, whose adaptation to the ETC
context could potentially unlock additional performance improvements.


\begin{acks}
  This work is partially supported by the European Research Council
  through the SENTIENT project (ERC-2017-STG \#755953).
\end{acks}

\bibliographystyle{ACM-Reference-Format}


\end{document}